\documentclass[english,nofootinbib, notitlepage, pra, english]{revtex4-1}
\usepackage[T1]{fontenc}
\usepackage[utf8]{inputenc}
\usepackage{geometry}
\usepackage{color}
\usepackage{babel}
\usepackage{amsmath}
\usepackage{amssymb}
\usepackage{graphicx}
\usepackage[unicode=true,pdfusetitle,
 bookmarks=true,bookmarksnumbered=false,bookmarksopen=false,
 breaklinks=false,pdfborder={0 0 0},pdfborderstyle={},backref=false,colorlinks=true]
 {hyperref}
\begin{document}
\global\long\def\ket#1{\left|#1\right\rangle }%

\global\long\def\bra#1{\left\langle #1\right|}%

\global\long\def\braket#1#2{\langle#1|#2\rangle}%

\global\long\def\triple#1#2#3{#1\cdot(#2\times#3)}%

\global\long\def\x{\times}%

\global\long\def\t{\cdot}%

\title{On the molecular information revealed by photoelectron angular distributions
  of isotropic samples}
\author{Andres F. Ordonez}
\email{ordonez@mbi-berlin.de}
\affiliation{Max-Born-Institut, Berlin, Germany}
\affiliation{Technische Universit\"at Berlin, Berlin, Germany}
\author{Olga Smirnova}
\email{smirnova@mbi-berlin.de}
\affiliation{Max-Born-Institut, Berlin, Germany}
\affiliation{Technische Universit\"at Berlin, Berlin, Germany}

\begin{abstract}
We propose an alternative approach to the description and analysis
of photoelectron angular distributions (PADs) resulting from isotropic
samples in the case of few-photon absorption via electric fields of
arbitrary polarization. As we demonstrate for the one- and two-photon
cases, this approach reveals the molecular frame information encoded
in the $b_{l,m}$ expansion coefficients of the PAD in a particularly
clear way. Our approach does not rely on explicit partial wave expansions
of the scattering wave function and the expressions we obtain are
therefore interpreted in terms of the vector field structure of the
photoionization dipole $\vec{D}(\vec{k})$ as a function of the photoelectron
momentum $\vec{k}$. This provides very compact expressions that
reveal how molecular rotational invariants couple to the setup (electric
field polarization and detectors) rotational invariants. We rely
heavily on this approach in a companion paper on tensorial chiral
setups. Here we apply this approach to one-photon ionization and find
that while $b_{0,0}$ depends only on the magnitude of $\vec{D}(\vec{k})$,
$b_{1,0}$ (non-zero for chiral molecules) is sensitive only to the
components of $\vec{D}(\vec{k})$ perpendicular to $\vec{k}$ encoded
in the propensity field $\vec{B}(\vec{k})\equiv i\vec{D}^{*}(\vec{k})\times\vec{D}(\vec{k})$,
and $b_{2,0}$ is sensitive only to the the component of $\vec{D}(\vec{k})$
along $\vec{k}$. We also analyze the resonantly enhanced two-photon
case where we show that $b_{0,0}$ and $b_{1,0}$ can be written in
terms of an effectively stretched $\vec{D}(\vec{k})$, and that $b_{1,0}$
and $b_{3,0}$ reveal structural information of the field $\vec{B}(\vec{k})$
encoded in three of its vector spherical harmonic expansion coefficients.
\end{abstract}
\maketitle

\section{Introduction}

Photoelectrons provide an important window into the structure of matter.
In the case of gas phase molecules it is remarkable that part of that
structural information, which goes beyond the energy spectrum of the
molecule, is imprinted in the PAD even when the molecules are randomly
oriented in space. A paramount example of this is photoelectron circular
dichroism (PECD), where the opposite enantiomers of a chiral molecule
(sharing the same energy spectrum) yield markedly different PADs when
illuminated with circularly polarized light \cite{ritchie_theory_1976,powis_photoelectron_2008}.
Motivated by the importance of enantiomeric recognition for the chemical
industry and by the fact that it occurs already within the electric-dipole
approximation and yields very strong enantiosensitive signals, PECD
has been studied across a wide range of molecular species \cite{ulrich_giant_2008,nahon_valence_2015}
and photoionization regimes \cite{lux_circular_2012,garcia_vibrationally_2013,dreissigacker_photoelectron_2014,beaulieu_universality_2016,beaulieu_attosecond-resolved_2017,turchini_conformational_2017}.
 Crucially, the extension of PECD into the realm of multiphoton ionization
provides both access to time-resolved ultra-fast enantiosensitive
electronic dynamics \cite{beaulieu_attosecond-resolved_2017,comby_relaxation_2016,beaulieu_photoexcitation_2018,ordonez_generalized_2018-1}
and the means to control the enantiosensitive signal observed in the
PAD \cite{demekhin_photoelectron_2018,goetz_quantum_2019}. 

The interpretation of these and other exciting yet intricate phenomena
relies on the mathematical formulation available for the description
of PADs in molecules. A cornerstone of this formulation is the partial
wave expansion of the scattering wave function, which is normally
performed at the outset of any PAD derivation \cite{ritchie_theory_1976,powis_photoelectron_2008,tully_angular_1968,chandra_photoelectron_1987,reid_photoelectron_2003},
in preparation for the orientation averaging step. Here we show that
one can arrive to insightful orientation-averaged expressions for
the $b_{l,m}$ coefficients describing the PAD without invoking the
expansion of the scattering wave function. In fact, doing so reveals
interesting physics that would be otherwise obscured by the partial
wave expansion itself\footnote{The standard expressions can be recovered by subsequent replacement
of the partial wave expansion.}. We already took advantage of a restricted version of this approach
(valid only for the $b_{1,0}$ coefficient) in the analysis of one-
and two-photon ionization of chiral samples \cite{beaulieu_photoexcitation_2018,ordonez_generalized_2018-1,ordonez_propensity_2019}.
There it played a fundamental role in the interpretation of the phenomena
and in establishing connections to other enantiosensitive effects
occurring within the electric-dipole approximation \cite{giordmaine_nonlinear_1965,patterson_enantiomer-specific_2013,patterson_sensitive_2013}
as well as to a geometrical effect in solids \cite{yao_valley-dependent_2008}.

With this approach at our disposal we ask a simple question. What
is the meaning of the molecular information encoded in the orientation-averaged
$b_{l,m}$ coefficients? Can it be understood as something else besides
the complex interference of (potentially many $\sim15$ \cite{powis_photoelectron_2008,stener_density_2004,giardini_angle-resolved_2005,harding_sensitivity_2006,tommaso_conformational_2006})
partial waves with different phase shifts? Since the photoionization
is determined by the photoionization dipole $\vec{D}(\vec{k})\equiv\langle\vec{k}\vert\hat{\vec{d}}\vert0\rangle$
between the ground state and the scattering state, each of the $b_{l,m}$
coefficients should tell us something different about the structure
of $\vec{D}(\vec{k})$. Furthermore, since we are dealing with isotropic
samples only rotational invariants of the molecule and of the setup
(electric field polarization and detectors) can be part of the answer.
How are these coupled to each other? Here we provide a method to approach
these questions in general, and provide concrete and perhaps surprisingly
simple answers for the case of all $b_{l,m}$ coefficients in one-photon
ionization with arbitrary polarization and for the coefficients $b_{0,0},$
$b_{1,0}$, and $b_{3,0}$ in two-photon ionization with circularly
polarized light. We will also make extensive use of this approach
in a companion paper \cite{ordonez_tensorial} dealing with tensorial
chiral setups and a novel type of enantiosensitive asymmetries in
PADs \cite{demekhin_photoelectron_2018,demekhin_photoelectron_2019}.

The paper is organized as follows: in Sec. \ref{sec:General-methodology}
we present the main derivation for the $b_{l,m}$ coefficients in
multiphoton ionization with fields of arbitrary polarization. Section
\ref{sec:one-photon} contains the analysis of one-photon ionization
and Sec. \ref{sec:PECD_two-photon} the analysis of two-photon resonantly-enhanced
ionization. Section \ref{sec:Conclusions} summarizes the conclusions
of this work. 

\section{General methodology\label{sec:General-methodology}}

The photoionization of an isotropic molecular sample results in a
photoelectron spectrum $W^{\mathrm{L}}(\vec{k}^{\mathrm{L}})$ given
by

\begin{equation}
W^{\mathrm{L}}(\vec{k}^{\mathrm{L}})=\int\mathrm{d}\varrho\,W^{\mathrm{L}}(\vec{k}^{\mathrm{L}},\varrho)
\end{equation}

where $W^{\mathrm{L}}(\vec{k}^{\mathrm{L}},\varrho)$ is the photoelectron
spectrum for a given molecular orientation $\varrho\equiv\alpha\beta\gamma$,
$\alpha\beta\gamma$ are the Euler angles, $\int\mathrm{d}\varrho\equiv\frac{1}{8\pi^{2}}\int_{0}^{2\pi}\mathrm{d}\alpha\int_{0}^{\pi}\mathrm{d}\beta\int_{0}^{2\pi}\mathrm{d}\gamma$
is the integral over all orientations, and the superscript $\mathrm{L}$
indicates vectors and functions in the laboratory frame. Since we
can always expand $W^{\mathrm{L}}(\vec{k}^{\mathrm{L}})$ into real
spherical harmonics\footnote{We will use tildes to distinguish the real spherical harmonics $\tilde{Y}_{l,m}$
from the usual complex spherical harmonics $Y_{l,m}$. See Appendix.
For $m=0$ we will omit the tilde.} $\tilde{Y}_{l}^{m}(\hat{k}^{\mathrm{L}})$,

\begin{align}
W^{\mathrm{L}}(\vec{k}^{\mathrm{L}}) & =\sum_{l,m}\tilde{b}_{l,m}\left(k\right)\tilde{Y}_{l}^{m}(\hat{k}^{\mathrm{L}}),\label{eq:W_expansion}
\end{align}

then any information about the molecule and the ionizing field encoded
in the photoelectron spectrum $W^{\mathrm{L}}(\vec{k}^{\mathrm{L}})$
is now neatly summarized in the expansion coefficients $\tilde{b}_{l,m}\left(k\right)$,

\begin{align}
\tilde{b}_{l,m}\left(k\right) & =\int\mathrm{d}\Omega_{k}^{\mathrm{L}}\tilde{Y}_{l}^{m}(\hat{k}^{\mathrm{L}})W^{\mathrm{L}}(\vec{k}^{\mathrm{L}}),\nonumber \\
 & =\int\mathrm{d}\Omega_{k}^{\mathrm{L}}\tilde{Y}_{l}^{m}(\hat{k}^{\mathrm{L}})\int\mathrm{d}\varrho\,W^{\mathrm{L}}(\vec{k}^{\mathrm{L}},\varrho),\label{eq:b_lm_trivial}
\end{align}

where $\int\mathrm{d}\Omega_{k}^{\mathrm{L}}\equiv\int_{0}^{\pi}\mathrm{d}\theta_{k}^{\mathrm{L}}\int_{0}^{2\pi}\mathrm{d}\phi_{k}^{\mathrm{L}}$,
$\hat{k}^{\mathrm{L}}=\left(1,\theta_{k}^{\mathrm{L}},\phi_{k}^{\mathrm{L}}\right)$
in spherical coordinates, and $\vec{k}^{\mathrm{L}}=k\hat{k}^{\mathrm{L}}$.
By the definition of a rotated function (see e.g. \cite{brink_angular_1968})
we have that the photoelectron spectrum in the molecular frame is
given by the relation $W^{\mathrm{M}}(\vec{k}^{\mathrm{M}},\varrho)=W^{\mathrm{L}}(\vec{k}^{\mathrm{L}},\varrho),$
or equivalently $W^{\mathrm{M}}(S^{-1}\left(\varrho\right)\vec{k}^{\mathrm{L}},\varrho)=W^{\mathrm{L}}(\vec{k}^{\mathrm{L}},\varrho)$,
where $\vec{k}^{\mathrm{L}}=S\left(\varrho\right)\vec{k}^{\mathrm{M}}$,
$S\left(\varrho\right)$ is the rotation matrix that takes vectors
from the molecular to the laboratory frame and the superscript $\mathrm{M}$
indicates vectors and functions in the molecular frame. This means
that 

\begin{align}
\tilde{b}_{l,m}\left(k\right) & =\int\mathrm{d}\Omega_{k}^{\mathrm{L}}\tilde{Y}_{l}^{m}(\hat{k}^{\mathrm{L}})\int\mathrm{d}\varrho\,W^{\mathrm{M}}(S^{-1}\left(\varrho\right)\vec{k}^{\mathrm{L}},\varrho),\nonumber \\
 & =\int\mathrm{d}\varrho\int\mathrm{d}\Omega_{k}^{\mathrm{L}}\,\tilde{Y}_{l}^{m}(\hat{k}^{\mathrm{L}})W^{\mathrm{M}}(S^{-1}\left(\varrho\right)\vec{k}^{\mathrm{L}},\varrho),
\end{align}

where in the second line we exchanged the integration order because
we want to make the change of variables $\vec{k}^{\mathrm{M}}=S^{-1}\left(\varrho\right)\vec{k}^{\mathrm{L}}$,
which only exists inside the integral over orientations and yields

\begin{align}
\tilde{b}_{l,m}\left(k\right) & =\int\mathrm{d}\varrho\int\mathrm{d}\Omega_{k}^{\mathrm{M}}\,\tilde{Y}_{l}^{m}(S\left(\varrho\right)\hat{k}^{\mathrm{M}})W^{\mathrm{M}}(\vec{k}^{\mathrm{M}},\varrho),\nonumber \\
 & =\int\mathrm{d}\Omega_{k}^{\mathrm{M}}\int\mathrm{d}\varrho\,\tilde{Y}_{l}^{m}(S\left(\varrho\right)\hat{k}^{\mathrm{M}})W^{\mathrm{M}}(\vec{k}^{\mathrm{M}},\varrho).\label{eq:b_lm_puzzling}
\end{align}

where in the second line we exchanged the integration order again
because now $\vec{k}^{\mathrm{M}}$ is an integration variable independent
of $\varrho$. At this point two questions arise: Why would we want
to have the photoelectron momentum in the molecular frame instead
of having it in the laboratory frame, where the photoelectron is actually
measured? And why would we prefer to do the integral over orientations
in Eq. \eqref{eq:b_lm_puzzling} instead of the apparently simpler
integral over orientations in Eq. \eqref{eq:b_lm_trivial}? The answer
to both questions has to do with the form that the photoelectron spectrum
$W^{\mathrm{M}}(\vec{k}^{\mathrm{M}},\varrho)$ takes in the case
of perturbative ionization.

\begin{figure}
\begin{centering}
\includegraphics[scale=0.5]{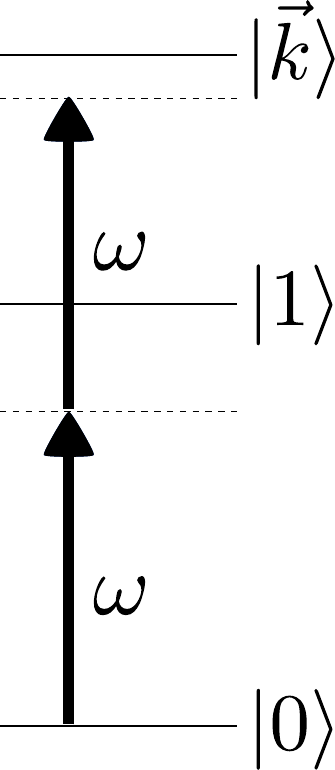}
\par\end{centering}
\centering{}\caption{Two-photon ionization through a bound state. \label{fig:Excitation-scheme}}
\end{figure}

As an example, let's consider the simple scenario depicted in Fig.
\ref{fig:Excitation-scheme}: two-photon absorption with a single
color field in a three-level system where the two lower levels are
bound and non-degenerate and the higher level is an infinitely degenerate
scattering state. For a Gaussian pulse with central frequency $\omega_{L}$
and spectral width $\gamma$ the field can be written as

\begin{eqnarray}
\vec{E}\left(\omega\right) & = & \sqrt{2\pi}\left[\frac{\vec{F}_{\omega_{L}}}{2}\delta_{\gamma}\left(\omega+\omega_{L}\right)+\frac{\vec{F}_{\omega_{L}}^{*}}{2}\delta_{\gamma}\left(\omega-\omega_{L}\right)\right],\qquad\delta_{\gamma}\left(\omega\right)\equiv\frac{e^{-\omega^{2}/\left(2\gamma^{2}\right)}}{\sqrt{2\pi\gamma^{2}}},\label{eq:field}
\end{eqnarray}

and the resulting second-order contribution to the probability amplitude
of the scattering state $\vert\vec{k}^{\mathrm{M}}\rangle$ reads
as

\begin{align}
a_{\vec{k}^{\mathrm{M}}}^{\left(2\right)}\left(\varrho\right) & =A^{\left(2\right)}\left(\vec{d}_{\vec{k}^{\mathrm{M}},1}^{\mathrm{L}}\cdot\vec{F}_{\omega_{L}}^{\mathrm{L}}\right)\left(\vec{d}_{1,0}^{\mathrm{L}}\cdot\vec{F}_{\omega_{L}}^{\mathrm{L}}\right)\label{eq:a2}
\end{align}

where $\vec{d}_{i,j}\equiv\langle i\vert\vec{d}\vert j\rangle$ is
the transition dipole matrix element and $A^{\left(2\right)}$ is
a function of the difference of the level spacings $\omega_{k1}-\omega_{10}$,
the total detuning $2\Delta=\omega_{k0}-2\omega_{L}$, and the spectral
width $\gamma$, $\omega_{ij}\equiv\omega_{i}-\omega_{j}$, $\omega_{i}$
is the energy of the state $\vert i\rangle$, and the superscript
$\left(2\right)$ indicates the order of the process. The photoelectron
spectrum in the molecular frame then reads as

\begin{align}
W^{\mathrm{M}}(\vec{k}^{\mathrm{M}},\varrho) & \equiv\left|a_{\vec{k}^{\mathrm{M}}}^{\left(2\right)}\right|^{2}=\left|A^{\left(2\right)}\right|^{2}\left(\vec{d}_{\vec{k}^{\mathrm{M}},1}^{\mathrm{L}*}\cdot\vec{F}_{\omega_{L}}^{\mathrm{L}*}\right)\left(\vec{d}_{1,0}^{\mathrm{L}*}\cdot\vec{F}_{\omega_{L}}^{\mathrm{L}*}\right)\left(\vec{d}_{\vec{k}^{\mathrm{M}},1}^{\mathrm{L}}\cdot\vec{F}_{\omega_{L}}^{\mathrm{L}}\right)\left(\vec{d}_{1,0}^{\mathrm{L}}\cdot\vec{F}_{\omega_{L}}^{\mathrm{L}}\right)
\end{align}

where the $\varrho$ dependence is implicit in the transition dipoles
according to $\vec{d}_{i,j}^{\mathrm{L}}=S\left(\varrho\right)\vec{d}_{i,j}^{\mathrm{M}}$.
Replacing in Eq. \eqref{eq:b_lm_puzzling} we obtain

\begin{equation}
\tilde{b}_{l,m}^{\left(2\right)}\left(k\right)=\left|A^{\left(2\right)}\right|^{2}\int\mathrm{d}\Omega_{k}^{\mathrm{M}}\int\mathrm{d}\varrho\,\tilde{Y}_{l}^{m}(\hat{k}^{\mathrm{L}})\left(\vec{d}_{\vec{k}^{\mathrm{M}},1}^{\mathrm{L}*}\cdot\vec{F}_{\omega_{L}}^{\mathrm{L}*}\right)\left(\vec{d}_{1,0}^{\mathrm{L}*}\cdot\vec{F}_{\omega_{L}}^{\mathrm{L}*}\right)\left(\vec{d}_{\vec{k}^{\mathrm{M}},1}^{\mathrm{L}}\cdot\vec{F}_{\omega_{L}}^{\mathrm{L}}\right)\left(\vec{d}_{1,0}^{\mathrm{L}}\cdot\vec{F}_{\omega_{L}}^{\mathrm{L}}\right).\label{eq:b_lm_tensor}
\end{equation}

Note that when written in component form, the product of four transition
dipole vectors (tensors of rank 1) in this expression forms irreducible
spherical tensors of rank $\mu$ up to $4$ (twice the number of photons
$N$ exchanged with the field) which transform according to the Wigner
matrix $\mathcal{D}^{\left(\mu\right)}\left(\varrho\right)$. Similarly,
the real spherical harmonic $\tilde{Y}_{l}^{m}(\hat{k}^{\mathrm{L}})$
is a superposition of two spherical tensors of rank $l$ that transform
according to $\mathcal{D}^{\left(l\right)}\left(\varrho\right)$ \cite{brink_angular_1968}.
Then, from Eq. \eqref{eq:b_lm_tensor} and the orthogonality relation
of the Wigner matrices \cite{brink_angular_1968} it is evident that
the $\tilde{b}_{l,m}^{\left(2\right)}$ coefficients with $l>l_{\mathrm{max}}=4$
(in general $l_{\mathrm{max}}=2N$) vanish, as is well known. Expressions
analogous to Eq. (\ref{eq:b_lm_tensor}) can be obtained for the case
of fields with multiple frequencies. In the case of terms $\tilde{b}_{l,m}^{\left(N_{1},N_{2}\right)}$
resulting from the interference of pathways involving $N_{1}$ and
$N_{2}$ photons we get $l_{\mathrm{max}}=N_{1}+N_{2}$.

\begin{figure}
\begin{centering}
\includegraphics[scale=0.3]{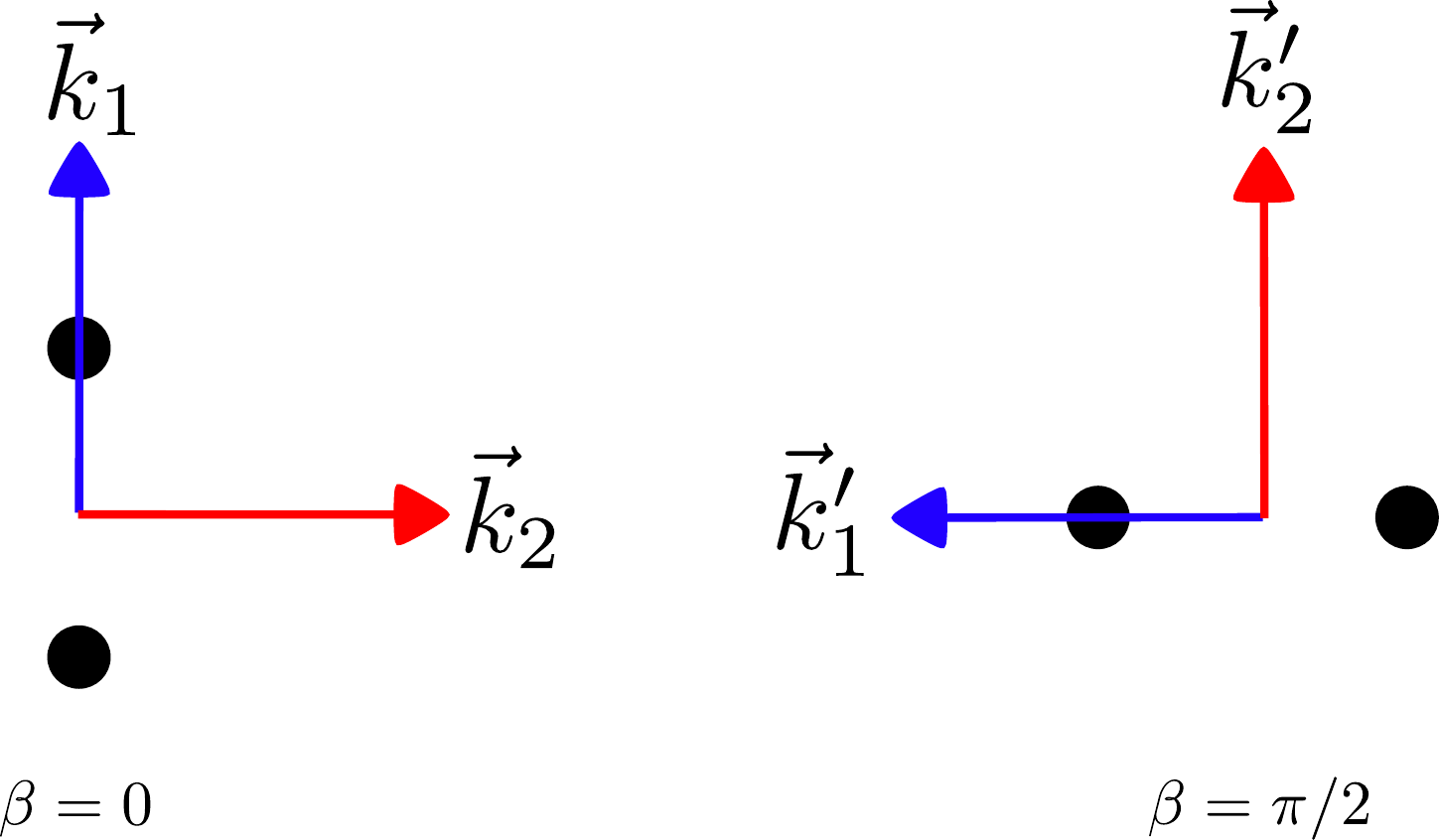}
\par\end{centering}
\caption{Two orientations of a diatomic molecule. The black circles indicate
the nuclei. The scattering state depends on the relative angle between
the molecular axis and the propagation direction $\vec{k}$ of the
outgoing (asymptotically plane) wave. Therefore, while the states
$\vert\vec{k}_{1}\rangle$ and $\vert\vec{k}_{1}^{\prime}\rangle$
satisfying $\vec{k}_{1}^{\mathrm{M}}=\vec{k}_{1}^{\prime\mathrm{M}}$
are related to each other by a simple rotation, the states $\vert\vec{k}_{1}\rangle$
and $\vert\vec{k}_{2}^{\prime}\rangle$ satisfying $\vec{k}_{1}^{\mathrm{L}}=\vec{k}_{2}^{\prime\mathrm{L}}$
are not related to each other in any simple way. \label{fig:Two-orientations}}

\end{figure}

If instead of relying on the Wigner matrices to perform the orientation
averaging we take into account that the spherical harmonics in Eq.
\eqref{eq:b_lm_tensor} are just polynomials of $k_{x}^{\mathrm{L}}/k$,
$k_{y}^{\mathrm{L}}/k$, and $k_{z}^{\mathrm{L}}/k$, then $\tilde{b}_{l,m}$
becomes a sum of terms of the form

\begin{multline}
\int\mathrm{d}\Omega_{k}^{\mathrm{M}}\int\mathrm{d}\varrho\,\left(\hat{k}^{\mathrm{L}}\cdot\hat{x}^{\mathrm{L}}\right)^{p}\left(\hat{k}^{\mathrm{L}}\cdot\hat{y}^{\mathrm{L}}\right)^{q}\left(\hat{k}^{\mathrm{L}}\cdot\hat{z}^{\mathrm{L}}\right)^{r}\\
\times\left(\vec{d}_{\vec{k}^{\mathrm{M}},1}^{\mathrm{L}*}\cdot\vec{F}_{\omega_{L}}^{\mathrm{L}*}\right)\left(\vec{d}_{1,0}^{\mathrm{L}*}\cdot\vec{F}_{\omega_{L}}^{\mathrm{L}*}\right)\left(\vec{d}_{\vec{k}^{\mathrm{M}},1}^{\mathrm{L}}\cdot\vec{F}_{\omega_{L}}^{\mathrm{L}}\right)\left(\vec{d}_{1,0}^{\mathrm{L}}\cdot\vec{F}_{\omega_{L}}^{\mathrm{L}}\right),\label{eq:b_lm_basic}
\end{multline}

where $p+q+r\leq l$ and $p+q+r$ has the same parity as $l$. The
vectors in this expression are of two types. The set \{$\hat{x}^{\mathrm{L}}$,
$\hat{y}^{\mathrm{L}}$, $\hat{z}^{\mathrm{L}}$, $\vec{F}_{\omega_{L}}^{\mathrm{L}}$\}
is fixed in the laboratory frame, while the set \{$\hat{k}^{\mathrm{M}}$,
$\vec{d}_{\vec{k}^{\mathrm{M}},1}^{\mathrm{M}}$, $\vec{d}_{1,0}^{\mathrm{M}}$\}
{[}which appears in the expression above rotated into the laboratory
frame $\vec{v}^{\mathrm{L}}=S\left(\varrho\right)\vec{v}^{\mathrm{M}}${]}
is fixed in the molecular frame. We take $\hat{k}^{\mathrm{M}}$ fixed
in the molecular frame because $\hat{k}^{\mathrm{M}}$ is the quantum
label that characterizes the scattering state $\vert\vec{k}^{\mathrm{M}}\rangle$,
which (like the bound states) is fixed to the molecular frame (see
e.g. Fig. \ref{fig:Two-orientations}). Equation \eqref{eq:b_lm_basic}
has the form we wanted to achieve, it is a product of scalar products
between vectors fixed in the molecular frame and vectors fixed in
the laboratory frame. In this form the integration over orientations
can be performed at once applying the technique in Ref. \cite{andrews_threedimensional_1977},
which yields a result of the form $\sum_{i}g_{i}M_{ij}f_{j}$, where
the $g_{i}$ are rotational invariants formed with the set of vectors
fixed in the molecular frame, the $f_{i}$ are rotational invariants
formed by the set of vectors fixed in the laboratory frame, and the
$M_{ij}$ are constants. Examples of such invariants will be given
in the next section. 

The structure of the rotational invariants and the fact that Eq. (\ref{eq:b_lm_basic})
involves only polar vectors allows us to conclude that if the number
of dot products in Eq. (\ref{eq:b_lm_basic}) is odd (even) then the
rotational invariants are pseudoscalars (scalars). This means that
enantiosensitivity can only be observed in coefficients $\tilde{b}_{l,m}^{\left(N\right)}$
such that $l$ is odd, in agreement with previous works (see e.g.
Refs. \cite{lux_circular_2012,rafiee_fanood_chiral_2014}). More interestingly,
for coefficients $\tilde{b}_{l,m}^{\left(N_{1},N_{2}\right)}$ resulting
from interference between pathways with $N_{1}$ and $N_{2}$ photons
the condition for enantiosensitivity is that $l+N_{1}+N_{2}$ is odd,
in agreement with the recent works in Refs. \cite{demekhin_photoelectron_2018,demekhin_photoelectron_2019}.
This is a general condition independent of the polarization of the
field and of the photon energies and will be explored in more detail
in the companion paper \cite{ordonez_tensorial}.

Now we will discuss two elementary applications of our methodology.
First, we will derive the expression for the $b_{l,m}^{\left(1\right)}$
coefficients in one-photon ionization and discuss the molecular information
they reveal. Afterwards we will derive and discuss the expressions
for the $b_{0,0}^{\left(2\right)}$, $b_{1,0}^{\left(2\right)}$,
and $b_{3,0}^{\left(2\right)}$ coefficients relevant for PECD in
two-photon ionization. Note that the expressions for $b_{0,0}^{\left(1\right)}$
and $b_{1,0}^{\left(1\right)}$ coefficients in one-photon ionization
and the $b_{1,0}^{\left(2\right)}$ coefficient in two-photon ionization
have already been derived using a less general procedure in Ref. \cite{ordonez_generalized_2018-1}.

\section{The photoionization dipole field and the $b_{l,m}$ coefficients
in one-photon ionization\label{sec:one-photon}}

For the field in Eq. \eqref{eq:field}, the first-order amplitude
of the scattering state $\vert\vec{k}^{\mathrm{M}}\rangle$ reads
as

\begin{equation}
a_{\vec{k}^{\mathrm{M}}}^{\left(1\right)}=A^{\left(1\right)}\left(\vec{D}^{\mathrm{L}}\cdot\vec{F}_{\omega_{L}}^{\mathrm{L}}\right),\qquad A^{\left(1\right)}=i\pi\delta_{\gamma}\left(\omega+\omega_{L}\right)\label{eq:a1}
\end{equation}

where we use the shorthand notations $\vec{D}^{\mathrm{M}}\equiv\vec{d}_{\vec{k}^{\mathrm{M}},0}^{\mathrm{M}}$,
$\vec{F}^{\mathrm{L}}\equiv\vec{F}_{\omega_{L}}^{\mathrm{L}}$, and
as usual $\vec{D}^{\mathrm{L}}=S\left(\varrho\right)\vec{D}^{\mathrm{M}}$.
From here on, we must keep in mind that $\vec{D}^{\mathrm{M}}=\vec{D}^{\mathrm{M}}(\vec{k}^{\mathrm{M}})$
is a complex vector field that depends on $\vec{k}^{\mathrm{M}}$.
That is, for a fixed initial state $\vert0\rangle$, $\vec{D}^{\mathrm{M}}(\vec{k}^{\mathrm{M}})$
is a mapping from the space of real three-dimensional vectors $\vec{k}^{\mathrm{M}}\in\mathbb{R}^{3}$
to the space of complex three-dimensional vectors $\vec{D}^{\mathrm{M}}\in\mathbb{C}^{3}$.
From Eq. \eqref{eq:a1} it is clear that this complex vector field
fully determines the response of the molecule to the ionizing field
and therefore the coefficients $\tilde{b}_{l,m}^{\left(1\right)}$
must correspond to properties of this vector field. The question is:
which property of the photoionization vector field $\vec{D}^{\mathrm{M}}(\vec{k}^{\mathrm{M}})$
is reflected in a given $\tilde{b}_{l,m}^{\left(1\right)}$ coefficient?

Since for first order amplitudes all frequencies act separately and
the most general polarization of a single frequency is elliptical
then we will assume an electric field that is elliptically polarized
in the $xy$ plane with its major axis along either the $\hat{x}^{\mathrm{L}}$
or the $\hat{y}^{\mathrm{L}}$ axis. From symmetry it follows that
the only non-zero $\tilde{b}_{l,m}^{\left(1\right)}$ coefficients
are $b_{0,0}^{\left(1\right)}$, $b_{1,0}^{\left(1\right)}$, $b_{2,0}^{\left(1\right)}$,
$\tilde{b}_{2,2}^{\left(1\right)}$, and $\tilde{b}_{2,-2}^{\left(1\right)}$
(we omit the tilde for $m=0$). With the help of Eqs. \eqref{eq:b_lm_puzzling},
\eqref{eq:b_lm_basic}, and performing the orientation integrals according
to Ref. \cite{andrews_threedimensional_1977} we obtain\footnote{Note that the expressions \eqref{eq:b00}-\eqref{eq:b22} apply for
arbitrary polarization of the electric field (in particular for linear
polarization along $\hat{z}^{\mathrm{L}}$). The assumption that the
field is contained in the $xy$ plane with its major axis along $\hat{x}^{\mathrm{L}}$
or $\hat{y}^{\mathrm{L}}$ simply serves the purpose of reducing the
number of non-zero $\tilde{b}_{l,m}^{\left(1\right)}$ coefficients.} (see Appendix)

\begin{equation}
b_{0,0}^{\left(1\right)}\left(k\right)=\left|A^{\left(1\right)}\right|^{2}\left\{ \frac{1}{3\sqrt{4\pi}}\int\mathrm{d}\Omega_{k}^{\mathrm{M}}\left|\vec{D}^{\mathrm{M}}\right|^{2}\right\} \left\{ \left|\vec{F}^{\mathrm{L}}\right|^{2}\right\} ,\label{eq:b00}
\end{equation}

\begin{equation}
b_{1,0}^{\left(1\right)}\left(k\right)=\left|A^{\left(1\right)}\right|^{2}\left\{ \frac{1}{6}\sqrt{\frac{3}{4\pi}}\int\mathrm{d}\Omega_{k}^{\mathrm{M}}\left[\hat{k}^{\mathrm{M}}\cdot\left(\vec{D}^{\mathrm{M}*}\times\vec{D}^{\mathrm{M}}\right)\right]\right\} \left\{ \hat{z}^{\mathrm{L}}\cdot\left(\vec{F}^{\mathrm{L}*}\times\vec{F}^{\mathrm{L}}\right)\right\} ,\label{eq:b10}
\end{equation}

\begin{equation}
b_{2,0}^{\left(1\right)}\left(k\right)=\left|A^{\left(1\right)}\right|^{2}\left\{ \frac{1}{12\sqrt{5\pi}}\int\mathrm{d}\Omega_{k}^{\mathrm{M}}\left(3\left|\hat{k}^{\mathrm{M}}\cdot\vec{D}^{\mathrm{M}}\right|^{2}-\left|\vec{D}^{\mathrm{M}}\right|^{2}\right)\right\} \left\{ 3\left|\hat{z}^{\mathrm{L}}\cdot\vec{F}^{\mathrm{L}}\right|^{2}-\left|\vec{F}^{\mathrm{L}}\right|^{2}\right\} ,\label{eq:b20}
\end{equation}

\begin{equation}
\tilde{b}_{2,2}^{\left(1\right)}\left(k\right)=\left|A^{\left(1\right)}\right|^{2}\left\{ \frac{1}{4\sqrt{15\pi}}\int\mathrm{d}\Omega_{k}^{\mathrm{M}}\left(3\left|\hat{k}^{\mathrm{M}}\cdot\vec{D}^{\mathrm{M}}\right|^{2}-\left|\vec{D}^{\mathrm{M}}\right|^{2}\right)\right\} \left\{ \left|\hat{x}^{\mathrm{L}}\cdot\vec{F}^{\mathrm{L}}\right|^{2}-\left|\hat{y}^{\mathrm{L}}\cdot\vec{F}^{\mathrm{L}}\right|^{2}\right\} ,\label{eq:b22}
\end{equation}

and $\tilde{b}_{2,-2}^{\left(1\right)}=0$, a peculiarity of the one-photon
case. That is, each $\tilde{b}_{l,m}^{\left(1\right)}$ coefficient
is the product of: a coupling term $\left|A^{\left(1\right)}\right|^{2}$
depending on the energy level spacing of the molecule and the spectrum
of the electric field, a molecular term expressed in the molecular
frame and averaged over all $\hat{k}^{\mathrm{M}}$ directions, and
a setup (field and laboratory axes) term expressed in the laboratory
frame. Unlike the usual expressions for $\tilde{b}_{l,m}^{\left(1\right)}$
(see e.g. \cite{ritchie_theory_1976}), Eqs. \eqref{eq:b00}-\eqref{eq:b22}
provide a rather simple expression for the molecular terms which,
as we will now discuss, are simply related to concrete properties
of the photoionization vector field $\vec{D}^{\mathrm{M}}$.

As expected, equation \eqref{eq:b00} shows that $b_{0,0}^{\left(1\right)}$,
which is simply the total cross section, records only the $\hat{k}^{\mathrm{M}}$-averaged
value of the magnitude of the field $\vec{D}^{\mathrm{M}}(\vec{k}^{\mathrm{M}})$.
More interestingly, Eq. \eqref{eq:b10} shows that $b_{1,0}^{\left(1\right)}$
is sensitive to the $\hat{k}^{\mathrm{M}}$-averaged value of the
triple product $\hat{k}^{\mathrm{M}}\cdot(\vec{D}^{\mathrm{M}*}\times\vec{D}^{\mathrm{M}})$,
which, unlike $b_{0,0}^{\left(1\right)}$, depends on the angles between
$\vec{k}^{\mathrm{M}}$, $\vec{D}^{\mathrm{M}}$, and $\vec{D}^{\mathrm{M}*}$.
The meaning of this quantity can be made evident if we use an appropriate
basis for our vector field $\vec{D}^{\mathrm{M}}$. Starting from
the unit vectors in spherical coordinates $\hat{k}$, $\hat{\theta}_{k}$,
and $\hat{\varphi}_{k}$ we define spherical vectors

\begin{equation}
\hat{k}_{\pm}^{\mathrm{M}}=\mp\frac{\hat{\theta}_{k}\mp i\hat{\varphi}_{k}}{\sqrt{2}}.\label{eq:spherical_radial}
\end{equation}

If we now write $\vec{D}^{\mathrm{M}}$ in terms of these \emph{contravariant
helicity-basis} vectors \cite{varschalovich_quantum_1988},

\begin{equation}
\vec{D}^{\mathrm{M}}=D_{+}^{\mathrm{M}}\hat{k}_{+}^{\mathrm{M}}+D_{-}^{\mathrm{M}}\hat{k}_{-}^{\mathrm{M}}+D_{k}^{\mathrm{M}}\hat{k}^{\mathrm{M}},\label{eq:d_spherical}
\end{equation}

then 

\begin{equation}
\hat{k}^{\mathrm{M}}\cdot\left(i\vec{D}^{\mathrm{M}*}\times\vec{D}^{\mathrm{M}}\right)=\left|D_{+}^{\mathrm{M}}\right|^{2}-\left|D_{-}^{\mathrm{M}}\right|^{2}.\label{eq:b10_dichroism}
\end{equation}

\begin{figure}
\begin{centering}
\includegraphics[scale=0.3]{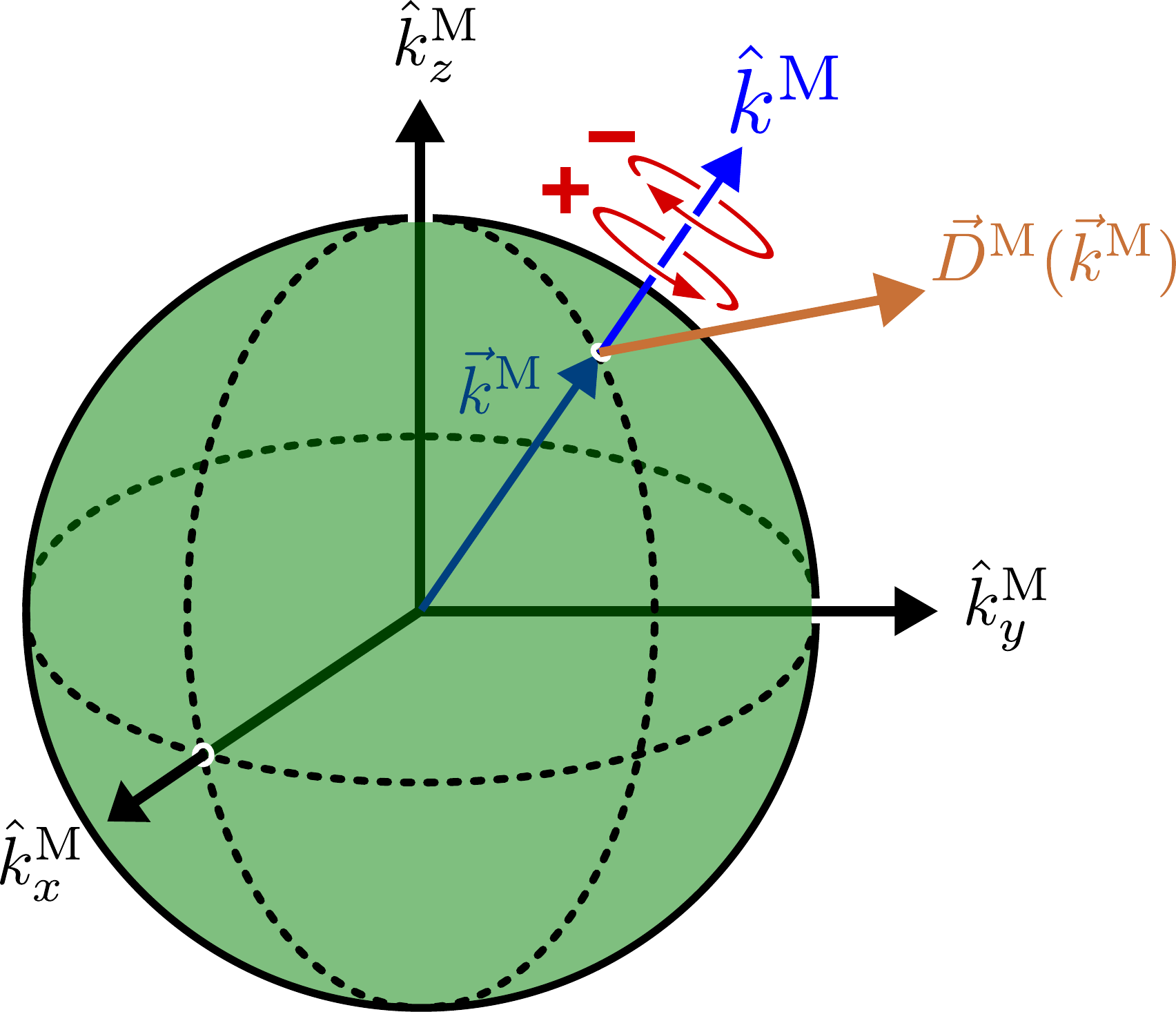}
\par\end{centering}
\caption{Sketch of the photoionization dipole $\vec{D}^{\mathrm{M}}(\vec{k}^{\mathrm{M}})\equiv\langle\vec{k}^{\mathrm{M}}\vert\vec{d}\vert0\rangle$
for a particular value of the photoelectron momentum $\vec{k}^{\mathrm{M}}$.
Red circular arrows indicate the direction of left ($+$) and right
($-$) circular polarization with respect to $\hat{k}^{\mathrm{M}}$.
\label{fig:circular_k}}

\end{figure}

The right hand side of Eq. \eqref{eq:b10_dichroism} is analogous
to the $s_{3}$ Stokes parameter for light waves in the circular polarization
basis, which describes the difference in intensity between left and
right circular polarization \cite{john_david_jackson_classical_nodate}.
Here we identify the right hand side of Eq. \eqref{eq:b10_dichroism}
with the circular dichroism (CD) of the photoionization vector field
$\vec{D}^{\mathrm{M}}$ in the direction $\hat{k}^{\mathrm{M}}$.
Indeed, the right hand side of Eq. \eqref{eq:b10_dichroism} is proportional
to the difference between the probability of inducing the transition
$\ket 0\rightarrow\vert\vec{k}^{\mathrm{M}}\rangle$ using left and
right circularly polarized light such that left ($+$) and right ($-$)
rotations are defined with respect to $\hat{k}^{\mathrm{M}}$ (see
Fig. \ref{fig:circular_k}). Therefore, the molecular term in $b_{1,0}^{\left(1\right)}$
is simply the $\hat{k}^{\mathrm{M}}$-averaged value of the $\vec{k}^{\mathrm{M}}$-specific
CD in the molecular frame. Note that the $\vec{k}^{\mathrm{M}}$-specific
CD can be non-zero even for achiral molecules, but its average over
$\hat{k}^{\mathrm{M}}$ is only non-zero for chiral molecules. Further
discussion of $b_{1,0}^{\left(1\right)}$ can be found in Ref. \cite{ordonez_propensity_2019}.
Remarkably, Eq. \eqref{eq:b10_dichroism} shows that $b_{1,0}^{\left(1\right)}$
depends only on the tangential components of $\vec{D}^{\mathrm{M}}$,
namely $D_{+}^{\mathrm{M}}$ and $D_{-}^{\mathrm{M}}$ (or equivalently
$D_{\theta}^{\mathrm{M}}$ and $D_{\varphi}^{\mathrm{M}}$). Since
the electric field term of $b_{1,0}^{\left(1\right)}$ has the same
form as the molecular part we can apply a similar procedure and rewrite
$b_{1,0}^{\left(1\right)}$ as

\begin{equation}
b_{1,0}^{\left(1\right)}=\left|A^{\left(1\right)}\right|^{2}\left\{ \frac{1}{6}\sqrt{\frac{3}{4\pi}}\int\mathrm{d}\Omega_{k}^{\mathrm{M}}\left(\left|D_{+}^{\mathrm{M}}\right|^{2}-\left|D_{-}^{\mathrm{M}}\right|^{2}\right)\right\} \left\{ \left|F_{+}^{\mathrm{L}}\right|^{2}-\left|F_{-}^{\mathrm{L}}\right|^{2}\right\} .\label{eq:b10_alt}
\end{equation}

where

\begin{equation}
\vec{F}^{\mathrm{L}}=F_{+}^{\mathrm{L}}\hat{\epsilon}_{+}^{\mathrm{L}}+F_{-}^{\mathrm{L}}\hat{\epsilon}_{-}^{\mathrm{L}}+F_{0}^{\mathrm{L}}\hat{z}^{\mathrm{L}},\label{eq:F_spherical}
\end{equation}

\begin{equation}
\hat{\epsilon}_{\pm}^{\mathrm{L}}=\frac{\hat{x}^{\mathrm{L}}\pm i\hat{y}^{\mathrm{L}}}{\sqrt{2}},\label{eq:spherical_cartesian}
\end{equation}

and $\left|F_{+}^{\mathrm{L}}\right|^{2}-\left|F_{-}^{\mathrm{L}}\right|^{2}$
is the Stokes parameter $s_{3}$ in the circular polarization basis
\cite{john_david_jackson_classical_nodate}.

If we now take the ratio between $b_{1,0}^{\left(1\right)}$ {[}Eq.
\eqref{eq:b10_alt}{]} and $b_{0,0}^{\left(1\right)}$ {[}Eq. \eqref{eq:b00}{]}
we get rid of the coupling term $\vert A^{\left(1\right)}\vert^{2}$,
and therefore obtain an expression which factorizes into a purely
molecular and a purely electric field part,

\begin{equation}
\beta_{1}^{\left(1\right)}\equiv\frac{\sqrt{3}b_{1,0}^{\left(1\right)}}{b_{0,0}^{\left(1\right)}}=\frac{3}{2}\left\{ \frac{\int\mathrm{d}\Omega_{k}^{\mathrm{M}}\left(\left|D_{+}^{\mathrm{M}}\right|^{2}-\left|D_{-}^{\mathrm{M}}\right|^{2}\right)}{\int\mathrm{d}\Omega_{k}^{\mathrm{M}}\left|\vec{D}^{\mathrm{M}}\right|^{2}}\right\} \left\{ \frac{\left|F_{+}^{\mathrm{L}}\right|^{2}-\left|F_{-}^{\mathrm{L}}\right|^{2}}{\left|\vec{F}^{\mathrm{L}}\right|^{2}}\right\} .\label{eq:beta_1}
\end{equation}

As discussed in Ref. \cite{ordonez_propensity_2019-1}, for any number
of photons $N$, we have that $\beta_{1}^{\left(N\right)}\equiv\sqrt{3}b_{1,0}^{\left(N\right)}/b_{0,0}^{\left(N\right)}=3j_{z}^{\left(N\right)}/j_{r}^{\left(N\right)}$,
where $j_{z}^{\left(N\right)}$ is the net photoelectron current (i.e.
vector sum of photoelectron currents in all directions) and $j_{r}^{\left(N\right)}$
is the total photoelectron current (i.e. sum of magnitudes of photoelectron
currents in all directions). Equation \eqref{eq:beta_1} shows that
the molecular factor is a measure of the degree of ``circular polarization''
of the photoionization vector field $\vec{D}^{\mathrm{M}}(\vec{k}^{\mathrm{M}})$
and takes values between $-1$ and $+1$, which correspond to the
limits $\vec{D}^{\mathrm{M}}=D_{+}^{\mathrm{M}}\hat{k}_{+}^{\mathrm{M}}$
(left circularly polarized $\vec{D}^{\mathrm{M}}$) and $\vec{D}^{\mathrm{M}}=D_{-}^{\mathrm{M}}\hat{k}_{-}^{\mathrm{M}}$
(right circularly polarized $\vec{D}^{\mathrm{M}}$), respectively.
Since the electric field factor is also a measure of the circular
polarization of the electric field, then $\beta_{1}^{\left(1\right)}=3j_{z}^{\left(1\right)}/j_{r}^{\left(1\right)}$
is given by the product of the $\hat{k}^{\mathrm{M}}$-averaged ``circular
polarization'' of the photoionization vector field $\vec{D}^{\mathrm{M}}(\vec{k}^{\mathrm{M}})$
and the circular polarization of the ionizing electric field. Clearly,
for a known electric field, $\beta_{1}^{\left(1\right)}$ is a measure
of the $\hat{k}^{\mathrm{M}}$-averaged ``circular polarization''
of $\vec{D}^{\mathrm{M}}(\vec{k}^{\mathrm{M}})$. 

\begin{figure}
\begin{centering}
\includegraphics[scale=0.35]{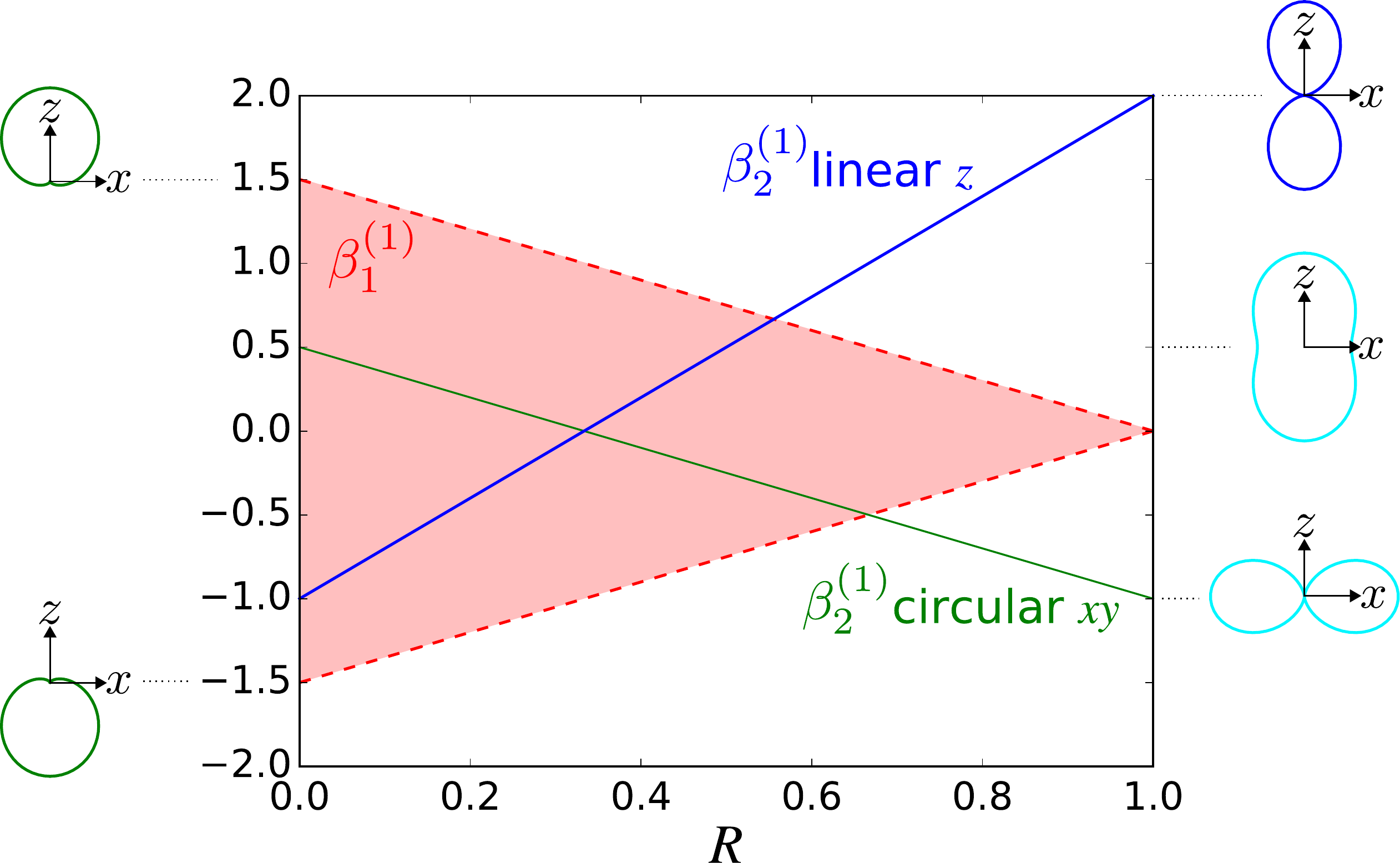}
\par\end{centering}
\caption{The relation between $\beta_{2}^{\left(1\right)}$ and the molecular
property $R$, which measures how radial the photoionization dipole
field $\vec{D}^{\mathrm{M}}(\vec{k}^{\mathrm{M}})=\langle\vec{k}^{\mathrm{M}}\vert\vec{d}^{\mathrm{M}}\vert0\rangle$
is in average for a given $k$ {[}see Eqs. \eqref{eq:beta_2} and
\eqref{eq:R}{]}, for the case of linear polarization along $\hat{z}$
(thick blue line) and circular polarization in the $\hat{x}\hat{y}$
plane (narrow green line). The red shaded area shows the range of
values that $\beta_{1}^{\left(1\right)}$ can take for a given value
of $\beta_{2}^{\left(1\right)}$ (and correspondingly of $R$) for
the circularly polarized case {[}see Eq. (\ref{eq:beta1_beta2_inequality}){]}.
$\beta_{1}^{\left(1\right)}$ is zero for linear polarization. The
insets on the left show the angular distributions for the extreme
values $\beta_{1}^{\left(1\right)}=\pm1.5$ and $\beta_{2}^{\left(2\right)}=0.5$
obtained for circular polarization. The insets on the right show the
angular distributions for $\beta_{1}^{\left(1\right)}=0$ (for simplicity)
and $\beta_{2}^{\left(1\right)}=-1$ (bottom), $\beta_{2}^{\left(1\right)}=0.5$
(center), and $\beta_{2}^{\left(1\right)}=2$ (top), which are the
values reached for linear and circular polarizations in the limits
$R=0$ (tangential $\vec{D}^{\mathrm{M}}$) and $R=1$ (radial $\vec{D}^{\mathrm{M}}$).
\label{fig:beta_R}}
\end{figure}

Moving on to the next coefficient, Eq. \eqref{eq:b20} shows that,
complementarily to $b_{0,0}^{\left(1\right)}$ and $b_{1,0}^{\left(1\right)}$
which depend on the magnitude and on the tangential components of
$\vec{D}^{\mathrm{M}}$, respectively, the coefficient $b_{2,0}^{\left(2\right)}$
depends on the projection of $\vec{D}^{\mathrm{M}}$ along $\hat{k}^{\mathrm{M}}$,
i.e. on its radial component $D_{k}^{\mathrm{M}}$ {[}see Eq. \eqref{eq:d_spherical}){]}.
We can also consider the ratio between $b_{2,0}^{\left(2\right)}$
and $b_{1,0}^{\left(2\right)}$ to get rid of the coupling term, and
obtain the asymmetry parameter\footnote{The factor of $\sqrt{3}$ in Eq. \eqref{eq:beta_1} and $\sqrt{5}$
in Eq. \eqref{eq:beta_1} are included to recover the ratio obtained
when the expansion is done in terms of Legendre polynomials (instead
spherical harmonics), as is usual for the cylindrically symmetric
cases when the light is either linearly polarized along $z$ or circularly
polarized in the $xy$ plane.},

\begin{equation}
\beta_{2}^{\left(1\right)}\equiv\frac{\sqrt{5}b_{2,0}^{\left(1\right)}\left(k\right)}{b_{0,0}^{\left(1\right)}\left(k\right)}=\frac{1}{2}\left\{ 3\frac{\int\mathrm{d}\Omega_{k}^{\mathrm{M}}\left|D_{k}^{\mathrm{M}}\right|^{2}}{\int\mathrm{d}\Omega_{k}^{\mathrm{M}}\left|\vec{D}^{\mathrm{M}}\right|^{2}}-1\right\} \left\{ 3\frac{\left|F_{z}^{\mathrm{L}}\right|^{2}}{\left|\vec{F}^{\mathrm{L}}\right|^{2}}-1\right\} .\label{eq:beta_2}
\end{equation}

which satisfies the well known fact that the values of $\beta_{2}^{\left(1\right)}$
for linear polarization along $z$ and circular polarization in the
$xy$ plane are related to each other by a factor of -2 \cite{reid_photoelectron_2003}.
More interestingly, we see that $\beta_{2}^{\left(1\right)}$ is a
linear function of the molecular property

\begin{equation}
R\equiv\frac{\int\mathrm{d}\Omega_{k}^{\mathrm{M}}\left|D_{k}^{\mathrm{M}}\right|^{2}}{\int\mathrm{d}\Omega_{k}^{\mathrm{M}}\left|\vec{D}^{\mathrm{M}}\right|^{2}},\qquad0\leq R\leq1,\label{eq:R}
\end{equation}

which measures to what extent the vector field $\vec{D}^{\mathrm{M}}(\vec{k}^{\mathrm{M}})$
is a radial field and takes values between $0$ and $1$, corresponding
to the limits $\vec{D}^{\mathrm{M}}=D_{+}^{\mathrm{M}}\hat{k}_{+}^{\mathrm{M}}+D_{-}^{\mathrm{M}}\hat{k}_{-}^{\mathrm{M}}$
(tangential field) and $\vec{D}^{\mathrm{M}}=D_{k}^{\mathrm{M}}\hat{k}$
(radial field), respectively. Figure \ref{fig:beta_R} shows $\beta_{2}^{\left(1\right)}$
as a function of $R$ for linear ($\vec{F}^{\mathrm{L}}=F_{z}\hat{z}^{\mathrm{L}}$)
and circular polarization ($\vec{F}^{\mathrm{L}}=F_{\pm}\hat{\epsilon}_{\pm}$
) along with the angular distributions obtained in the limits $R=0$
(tangential $\vec{D}^{\mathrm{M}}$) and $R=1$ (radial $\vec{D}^{\mathrm{M}}$).
We can see that for both linearly and circularly polarized fields,
a predominantly tangential field $\vec{D}^{\mathrm{M}}$ will yield
most photoelectrons with directions perpendicular to the electric
field, while a predominantly radial field $\vec{D}^{\mathrm{M}}$
will yield most photoelectrons with directions parallel to the electric
field. 

Figure \ref{fig:beta_R} also shows the range of values that $\beta_{1}^{\left(1\right)}$
can take as a function of $R$ for light circularly polarized in the
$xy$ plane. This follows from using the expressions for $b_{0,0}^{\left(1\right)}$,
$b_{1,0}^{\left(1\right)}$, and $b_{2,0}^{\left(1\right)}$ in Eqs.
(\ref{eq:b00}), (\ref{eq:b20}), and (\ref{eq:b10_alt}), and taking
into account that $\vert D_{+}\vert^{2}+\vert D_{-}\vert^{2}\geq\vert D_{+}\vert^{2}-\vert D_{-}\vert^{2}$,
one can show that for circularly polarized light $\beta_{1}^{\left(1\right)}$
and $\beta_{2}^{\left(2\right)}$ satisfy the inequality (see Appendix)

\begin{equation}
\vert\beta_{1}^{\left(1\right)}\vert\leq1+\beta_{2}^{\left(1\right)}.\label{eq:beta1_beta2_inequality}
\end{equation}

This inequality follows naturally from the fact that, for circularly
polarized light, small values of $\beta_{2}^{\left(1\right)}$ indicate
that the field $\vec{D}(\vec{k})$ is (in average) mostly radial and
therefore the tangential components along with $\beta_{1}^{\left(1\right)}$
are very small. On the contrary, big values of $\beta_{2}^{\left(1\right)}$
indicate that the field $\vec{D}(\vec{k})$ has (in average) a very
small radial component, which means that the field is mostly tangential
and can potentially display a large dichroism $\vert D_{+}\vert^{2}-\vert D_{-}\vert^{2}$.
The maximal value of $\vert\beta_{1}^{\left(1\right)}\vert=3/2$ and
occurs for $\beta_{2}^{\left(1\right)}=0.5$ {[}$\vec{D}(\vec{k})$
purely tangential{]}. As explained in Ref. \cite{ordonez_propensity_2019-1}
{[}Eqs. (9) and (10){]}, the net photoelectron current {[}i.e. the
vector sum of all photoelectron currents{]} is given by $j_{z}^{\mathrm{L}}(k)=\sqrt{4\pi/3}kb_{1,0}^{\left(1\right)}(k)$
and the total photoelectron current {[}i.e. the sum of the magnitudes
of all photoelectron currents{]} is given by $j_{r}\left(k\right)=\sqrt{4\pi}kb_{0,0}^{\left(1\right)}(k)$.
Therefore, the maximum value of the ratio of net photoelectron current
to total current is $\vert j_{z}\vert/j_{r}=1/2$. Note that Eq. \eqref{eq:beta1_beta2_inequality}
can also be derived exclusively from the condition that the angular
distribution $W^{\mathrm{L}}(\vec{k}^{\mathrm{L}})$ {[}Eq. (\ref{eq:W_expansion}){]}
is positive for every $\vec{k}^{\mathrm{L}}$.

Finally, Eq. \eqref{eq:b22} shows that, up to constants, $\tilde{b}_{2,2}^{\left(1\right)}$
differs from $b_{2,0}^{\left(1\right)}$ only in the electric field
factor, which in the case of $\tilde{b}_{2,2}^{\left(1\right)}$ yields
the $s_{1}$ Stokes parameter in the linear polarization basis \cite{john_david_jackson_classical_nodate}.
That is, $\tilde{b}_{2,2}^{\left(1\right)}$ and $b_{2,0}^{\left(1\right)}$
reveal the same information about the photoionization vector field
$\vec{D}^{\mathrm{M}}(\vec{k}^{\mathrm{M}})$ and differ only on the
electric field information they encode. This is a general property
of $\tilde{b}_{l,m}$ coefficients with the same value of $l$ and
corresponding to the same quantum pathway. It reflects the fact that
such coefficients differ only in their laboratory axes vectors {[}see
e.g. Eqs. \eqref{eq:b_lm_tensor} and \eqref{eq:b_lm_basic}{]} but
not on their molecular vectors (photoelectron momentum and transition
dipoles), and therefore they involve the same molecular rotational
invariants. 

\section{PECD in resonantly enhanced two-photon ionization\label{sec:PECD_two-photon}}

We begin by rewriting Eq. \eqref{eq:b_lm_tensor} as

\begin{equation}
\tilde{b}_{l,m}^{\left(2\right)}\left(k\right)=\frac{1}{2}\vert A^{\left(2\right)}\vert^{2}d^{2}\left|F\right|^{2}\int\mathrm{d}\Omega_{k}^{\mathrm{M}}\int\mathrm{d}\varrho\,\tilde{Y}_{l}^{m}(\hat{k}^{\mathrm{L}})\sin^{2}\beta\vert\vec{D}^{\mathrm{L}}\cdot\vec{F}^{\mathrm{L}}\vert^{2}.\label{eq:b_lm_2_circ}
\end{equation}

where we used the shorthand notation $\vec{D}^{\mathrm{M}}\equiv\vec{d}_{\vec{k}^{\mathrm{M}},1}^{\mathrm{M}}$
for the photoionization dipole from the intermediate state, $\vec{F}^{\mathrm{L}}\equiv\vec{F}_{\omega_{L}}^{\mathrm{L}}=F\left(1,i\sigma,0\right)/\sqrt{2}$,
$\sigma=\pm1$, and we chose the molecular axis so that $\vec{d}^{\mathrm{M}}\equiv\vec{d}_{1,0}^{\mathrm{M}}=d\hat{z}^{\mathrm{M}}$
and therefore $\vec{d}^{\mathrm{L}}=d(\sin\beta\cos\alpha,$$\sin\beta\sin\alpha$$,\cos\beta)$
where $\alpha\beta\gamma$ are the Euler angles in the $ZYZ$ convention,
and in particular $\beta$ is the angle between the molecular and
laboratory $\hat{z}$ axes. This yields $\vert\vec{d}^{\mathrm{L}}\cdot\vec{F}^{\mathrm{L}}\vert^{2}=\frac{1}{2}d^{2}\left|F\right|^{2}\sin^{2}\beta$.
Written like this, the second order coefficients $\tilde{b}_{l,m}^{\left(2\right)}\left(k\right)$
take the form of the first order coefficients $\tilde{b}_{l,m}^{\left(1\right)}$
for an anisotropic (in this case anti-aligned) sample with an orientation
distribution given by $w\left(\beta\right)\propto\sin^{2}\beta$ and
an initial state $\ket 1$ instead of $\ket 0$ (see also Refs. \cite{ordonez_propensity_2019,lehmann_imaging_2013,goetz_theoretical_2017}).
Such anisotropy gives a certain preference to the $z$ components
of the molecular vectors. Performing the orientation averaging according
to Ref. \cite{andrews_threedimensional_1977}, the expressions for
the total absorption $b_{0,0}^{\left(2\right)}$, and for the enantiosensitive
terms $b_{1,0}^{\left(2\right)}$ and $b_{3,0}^{\left(2\right)}$
yield\footnote{See also Ref. \cite{ordonez_generalized_2018-1}.}
(see Appendix)

\begin{align}
b_{0,0}^{\left(2\right)} & =C\left\{ \frac{1}{3}\frac{1}{\sqrt{4\pi}}\int\mathrm{d}\Omega_{k}^{\mathrm{M}}\vert D_{\mathrm{eff(0,0)}}^{\mathrm{M}}\vert^{2}\right\} ,\label{eq:b00_2}
\end{align}

\begin{align}
b_{1,0}^{\left(2\right)} & =\sigma C\left\{ \frac{1}{6}\sqrt{\frac{3}{4\pi}}\int\mathrm{d}\Omega_{k}^{\mathrm{M}}\left[\hat{k}^{\mathrm{M}}\cdot\left(i\vec{D}_{\mathrm{eff}\left(1,0\right)}^{\mathrm{M}*}\times\vec{D}_{\mathrm{eff}\left(1,0\right)}^{\mathrm{M}}\right)\right]\right\} ,\label{eq:b10_2}\\
 & =\sigma C\left\{ \frac{1}{6}\sqrt{\frac{3}{4\pi}}\int\mathrm{d}\Omega_{k}^{\mathrm{M}}\left[\vec{K}_{1,0}^{\mathrm{M}}\cdot\left(i\vec{D}^{\mathrm{M}*}\times\vec{D}^{\mathrm{M}}\right)\right]\right\} \frac{1}{5}\sqrt{\frac{17}{5}},\label{eq:b10_2_B}
\end{align}

\begin{align}
b_{3,0}^{\left(2\right)} & =\sigma C\left\{ \frac{1}{35\sqrt{3}}\sqrt{\frac{7}{16\pi}}\int\mathrm{d}\Omega_{k}^{\mathrm{M}}\left[\vec{K}_{3,0}^{\mathrm{M}}\cdot\left(i\vec{D}^{\mathrm{M}*}\times\vec{D}^{\mathrm{M}}\right)\right]\right\} ,\label{eq:b30_2}
\end{align}

where $C\equiv d^{2}\left|F\right|^{4}\vert A^{\left(2\right)}\vert^{2}$
is a common factor to all $b_{l,m}^{\left(2\right)}$ coefficients
that simply encodes the bound-bound transition and the second order
character of the process, and the expressions for $\vec{D}_{\mathrm{eff}\left(0,0\right)}^{\mathrm{M}}$,
$\vec{D}_{\mathrm{eff}\left(1,0\right)}^{\mathrm{M}}$, $\vec{K}_{1,0}^{\mathrm{M}}$,
and $\vec{K}_{3,0}^{\mathrm{M}}$ are given below. We wrote Eqs. (\ref{eq:b00_2})-(\ref{eq:b30_2})
so that we can draw a parallel to the corresponding Eqs. (\ref{eq:b00})
and (\ref{eq:b10}) in the one-photon case. Equations (\ref{eq:b00_2})
and (\ref{eq:b10_2}) show that we can recover the forms obtained
in the one-photon case if we introduce effectively stretched photoionization
dipoles given by

\begin{equation}
\vec{D}_{\mathrm{eff}\left(0,0\right)}^{\mathrm{M}}\equiv\sqrt{\frac{3}{10}}\left(D_{x}^{\mathrm{M}},D_{y}^{\mathrm{M}},\frac{2}{\sqrt{3}}D_{z}^{\mathrm{M}}\right),\label{eq:Deff00-1}
\end{equation}

and 

\begin{equation}
\vec{D}_{\mathrm{eff}\left(1,0\right)}^{\mathrm{M}}\equiv\frac{1}{\sqrt{5}}\left(D_{x}^{\mathrm{M}},D_{y}^{\mathrm{M}},2D_{z}^{\mathrm{M}}\right).\label{eq:Deff10-1}
\end{equation}

In view of the discussion in Sec. \ref{sec:one-photon}, Eq. (\ref{eq:b00_2})
shows that $b_{0,0}^{\left(2\right)}$ records the $\hat{k}^{\mathrm{M}}$-averaged
magnitude of an effective photoionization dipole $\vec{D}_{\mathrm{eff}\left(0,0\right)}^{\mathrm{M}}(\vec{k}^{\mathrm{M}})$.
Similarly, Eq. (\ref{eq:b10_2}) shows that $b_{1,0}^{\left(2\right)}$
records the ``circular polarization'' {[}see Eq. (\ref{eq:b10_dichroism}){]}
or equivalently the $\vec{k}^{\mathrm{M}}$-averaged value of the
$\vec{k}^{\mathrm{M}}$-specific CD of an effective photoionization
dipole $\vec{D}_{\mathrm{eff}\left(1,0\right)}^{\mathrm{M}}(\vec{k}^{\mathrm{M}})$.
Their ratio, $\beta_{1}^{\left(2\right)}\equiv\sqrt{3}b_{1,0}^{\left(2\right)}/b_{0,0}^{\left(2\right)}$,
can be interpreted as the average ``circular polarization'' of $\vec{D}_{\mathrm{eff}\left(1,0\right)}^{\mathrm{M}}$
normalized with respect to the average magnitude of $\vec{D}_{\mathrm{eff}\left(0,0\right)}^{\mathrm{M}}$.

In the case of $b_{3,0}^{\left(2\right)}$, quadratic terms in $k_{z}$
(see Appendix) hinder a straightforward interpretation of the integrand
in terms of a effectively stretched $\vec{D}^{\mathrm{M}}$. However,
like $b_{1,0}^{\left(1\right)}$ and $b_{1,0}^{\left(2\right)}$ {[}Eqs.
(\ref{eq:b10}) and (\ref{eq:b10_2_B}){]}, Eq. (\ref{eq:b30_2})
shows that $b_{3,0}^{\left(2\right)}$ depends on the photoionization
dipole $\vec{D}^{\mathrm{M}}(\vec{k}^{\mathrm{M}})$ only through
the $\vec{k}^{\mathrm{M}}$-dependent field 

\begin{equation}
\vec{B}^{\mathrm{M}}\equiv i\vec{D}^{\mathrm{M}*}\times\vec{D}^{\mathrm{M}},\label{eq:B}
\end{equation}

and we can therefore attempt an interpretation of $b_{3,0}^{\left(2\right)}$
in terms of $\vec{B}^{\mathrm{M}}$ directly. 

We have already found rigorous physical interpretations for the projections
$\hat{A}^{\mathrm{M}}\cdot\vec{B}^{\mathrm{M}}$ for $\hat{A}^{\mathrm{M}}=\hat{x}^{\mathrm{M}},\hat{y}^{\mathrm{M}},\hat{z}^{\mathrm{M}},\hat{k}^{\mathrm{M}}$
(see \cite{ordonez_propensity_2019} and Sec. \ref{sec:one-photon}).
In these cases we found that $\hat{A}^{\mathrm{M}}\cdot\vec{B}^{\mathrm{M}}$
yields the $\hat{k}^{\mathrm{M}}$-specific CD associated to the transition
$\ket 0\rightarrow\vert\vec{k}^{\mathrm{M}}\rangle$ for light circularly
polarized with respect to the axis $\hat{A}^{\mathrm{M}}$ (see Fig.
\ref{fig:circular_k}). In fact, this interpretation is valid for
an arbitrary $\hat{A}^{\mathrm{M}}$. To see this, note that for a
given $\hat{A}^{\mathrm{M}}$ one can always build $\hat{A}^{\mathrm{M}}$-dependent
unit vectors $\hat{e}_{\pm}^{\mathrm{M}}$ associated to positive
and negative rotations around $\hat{A}^{\mathrm{M}}$, write $\vec{D}^{\mathrm{M}}=D_{+}^{\mathrm{M}}\hat{e}_{+}^{\mathrm{M}}+D_{-}^{\mathrm{M}}\hat{e}_{-}^{\mathrm{M}}+D_{0}^{\mathrm{M}}\hat{A}^{\mathrm{M}}$
and obtain $\hat{A}^{\mathrm{M}}\cdot\vec{B}^{\mathrm{M}}=\vert\vec{D}_{+}^{\mathrm{M}}\vert^{2}-\vert\vec{D}_{-}^{\mathrm{M}}\vert^{2}$.
This scalar product is evidently maximized for $\hat{A}^{\mathrm{M}}=\hat{B}^{\mathrm{M}}$,
and therefore the direction of $\vec{B}^{\mathrm{M}}$ indicates the
axis with respect to which the $\hat{k}^{\mathrm{M}}$-specific CD
is maximal. The magnitude of $\vec{B}^{\mathrm{M}}$ is then the magnitude
of such maximal $\hat{k}^{\mathrm{M}}$-specific CD. Light circularly
polarized with respect to axes perpendicular to $\vec{B}^{\mathrm{M}}$
yield zero $\hat{k}^{\mathrm{M}}$-specific CD. 

While Eq. (\ref{eq:b10}) shows that $b_{1,0}^{\left(1\right)}$ involves
the projection of $\vec{B}^{\mathrm{M}}$ on the radial vector $\hat{k}^{\mathrm{M}}$,
Eqs. (\ref{eq:b10_2_B}) and (\ref{eq:b30_2}) show that $b_{1,0}^{\left(2\right)}$
and $b_{3,0}^{\left(2\right)}$ involve the projection of $\vec{B}^{\mathrm{M}}$
on the vector fields $\vec{K}_{1,0}^{\mathrm{M}}$ and $\vec{K}_{3,0}^{\mathrm{M}}$,
respectively, defined as (see Appendix)

\begin{equation}
\vec{K}_{1,0}^{\mathrm{M}}\equiv\sqrt{\frac{5}{17}}\left(2\hat{k}^{\mathrm{M}}-\frac{k_{z}^{\mathrm{M}}}{k}\hat{z}^{\mathrm{M}}\right),\label{eq:K10}
\end{equation}

\begin{equation}
\vec{K}_{3,0}^{\mathrm{M}}\equiv\frac{\sqrt{3}}{2}\left\{ \left[1-5\left(\frac{k_{z}^{\mathrm{M}}}{k}\right)^{2}\right]\hat{k}^{\mathrm{M}}+2\frac{k_{z}^{\mathrm{M}}}{k}\hat{z}^{\mathrm{M}}\right\} ,\label{eq:K30}
\end{equation}

\begin{figure}
\begin{centering}
\includegraphics[scale=0.5]{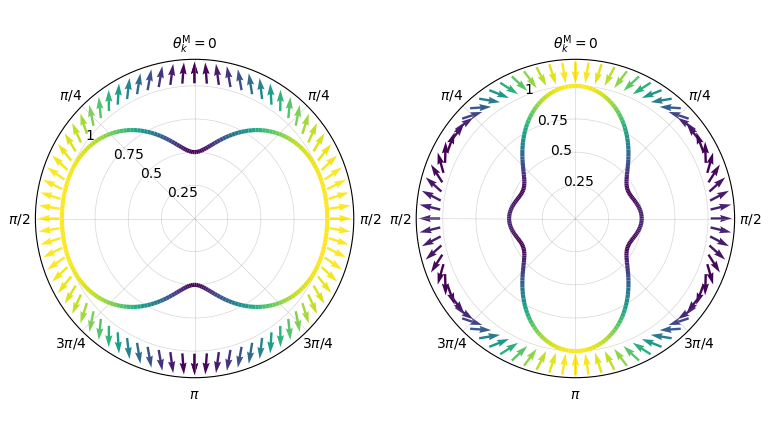}
\par\end{centering}
\caption{Direction (arrows) and magnitude (color and solid lines) of the vector
fields $\vec{K}_{1,0}^{\mathrm{M}}(\hat{k}^{\mathrm{M}})$ and $\vec{K}_{3,0}^{\mathrm{M}}(\hat{k}^{\mathrm{M}})$
in Eqs. (\ref{eq:K10}) and (\ref{eq:K30}).\label{fig:K10_K30}}

\end{figure}

and shown in Fig. \ref{fig:K10_K30} as a function of $\theta_{k}^{\mathrm{M}}$
on a plane parallel to $k_{z}^{\mathrm{M}}$. The integrations over
all $\hat{k}^{\mathrm{M}}$ directions in Eqs. (\ref{eq:b10_2}) and
(\ref{eq:b30_2}) tell us that $b_{1,0}^{\left(2\right)}$ and $b_{3,0}^{\left(2\right)}$
record the extent to which the vector field $\vec{B}^{\mathrm{M}}$
resembles the vector fields $\vec{K}_{1,0}^{\mathrm{M}}$ and $\vec{K}_{3,0}^{\mathrm{M}}$,
respectively, and therefore record structural information about $\vec{B}^{\mathrm{M}}$.
Such information can be made more explicit by expanding $\vec{K}_{1,0}^{\mathrm{M}}$,
$\vec{K}_{3,0}^{\mathrm{M}}$, and $\vec{B}^{\mathrm{M}}$ in terms
of vector spherical harmonics \cite{barrera_vector_1985} $\vec{Y}_{l,m}^{\mathrm{M}}(\hat{k}^{\mathrm{M}})\equiv Y_{l}^{m}(\hat{k}^{\mathrm{M}})\hat{k}^{\mathrm{M}}$,
$\vec{\Psi}_{l,m}^{\mathrm{M}}(\hat{k}^{\mathrm{M}})\equiv k\vec{\nabla}Y_{l,m}(\hat{k}^{\mathrm{M}})/\sqrt{l\left(l+1\right)}$,
and $\vec{\Phi}_{l,m}^{\mathrm{M}}(\hat{k}^{\mathrm{M}})\equiv\hat{k}^{\mathrm{M}}\times\vec{\Psi}_{l,m}^{\mathrm{M}}/\sqrt{l\left(l+1\right)}$,

\begin{equation}
\vec{K}_{1,0}^{\mathrm{M}}(\hat{k}^{\mathrm{M}})=\frac{2}{3}\sqrt{\frac{\pi}{17}}\left[5\sqrt{5}\vec{Y}_{0,0}^{\mathrm{M}}(\hat{k}^{\mathrm{M}})-2\vec{Y}_{2,0}^{\mathrm{M}}(\hat{k}^{\mathrm{M}})-\sqrt{6}\vec{\Psi}_{2,0}^{\mathrm{M}}(\hat{k}^{\mathrm{M}})\right],\label{eq:K10_expansion}
\end{equation}

\begin{equation}
\vec{K}_{3,0}^{\mathrm{M}}(\hat{k}^{\mathrm{M}})=2\sqrt{\frac{\pi}{5}}\left[-\sqrt{3}\vec{Y}_{2,0}^{\mathrm{M}}(\hat{k}^{\mathrm{M}})+\sqrt{2}\vec{\Psi}_{2,0}^{\mathrm{M}}(\hat{k}^{\mathrm{M}})\right],\label{eq:K30_expansion}
\end{equation}

\begin{equation}
\vec{B}^{\mathrm{M}}(\vec{k}^{\mathrm{M}})=\sum_{l,m}\left[\mathcal{B}_{l,m}^{Y}(k)\vec{Y}_{l,m}^{\mathrm{M}}(\hat{k}^{\mathrm{M}})+\mathcal{B}_{l,m}^{\Psi}(k)\vec{\Psi}_{l,m}^{\mathrm{M}}(\hat{k}^{\mathrm{M}})+\mathcal{B}_{l,m}^{\Phi}(k)\vec{\Phi}_{l,m}^{\mathrm{M}}(\hat{k}^{\mathrm{M}})\right].\label{eq:B_expansion-1}
\end{equation}

Replacing Eqs. (\ref{eq:B}), (\ref{eq:K10_expansion})-(\ref{eq:B_expansion-1})
in Eqs. (\ref{eq:b10_2_B}) and (\ref{eq:b30_2}) and using the orthonormality
relations for the vector spherical harmonics \cite{barrera_vector_1985},
we obtain 

\begin{equation}
b_{1,0}^{\left(2\right)}=\frac{\sigma C}{30\sqrt{15}}\left(5\sqrt{5}\mathcal{B}_{0,0}^{Y}-2\mathcal{B}_{2,0}^{Y}-\sqrt{6}\mathcal{B}_{2,0}^{\Psi}\right),\label{eq:b10_2_VSH}
\end{equation}

\begin{equation}
b_{3,0}^{\left(2\right)}=\frac{\sigma C}{10\sqrt{105}}\left(-\sqrt{3}\mathcal{B}_{2,0}^{Y}+\sqrt{2}\mathcal{B}_{2,0}^{\Psi}\right).\label{eq:b30_2_VSH}
\end{equation}

That is, while in the one-photon case $b_{1,0}^{\left(1\right)}$
encodes $\mathcal{B}_{0,0}^{Y}$ (because $\hat{k}^{\mathrm{M}}\propto\vec{Y}_{0,0}^{\mathrm{M}}$),
in the two-photon case $b_{1,0}^{\left(2\right)}$ and $b_{3,0}^{\left(2\right)}$
encode $\mathcal{B}_{0,0}^{Y}$, $\mathcal{B}_{2,0}^{Y}$, and $\mathcal{B}_{2,0}^{\Psi}$.
This motivates looking for a third linearly independent equation to
solve for $\mathcal{B}_{0,0}^{Y}$, $\mathcal{B}_{2,0}^{Y}$, and
$\mathcal{B}_{2,0}^{\Psi}$. This is delivered by the equation for
$b_{1,0}^{\prime\left(2\right)}$ for the complementary process where
the first photon is linearly polarized along $\hat{z}^{\mathrm{L}}$
and the second photon is circularly polarized in the $\hat{x}^{\mathrm{L}}\hat{y}^{\mathrm{L}}$
plane (see Appendix),

\begin{equation}
b_{1,0}^{\prime\left(2\right)}=\frac{\sigma C}{30\sqrt{15}}\left(5\sqrt{5}\mathcal{B}_{0,0}^{Y}+4\mathcal{B}_{2,0}^{Y}+2\sqrt{6}\mathcal{B}_{2,0}^{\Psi}\right).\label{eq:b10_2_lin}
\end{equation}

Equation (\ref{eq:b10_2_lin}) together with (\ref{eq:b10_2_VSH})
and (\ref{eq:b30_2_VSH}) yield

\begin{equation}
\mathcal{B}_{0,0}^{Y}=\frac{2\sqrt{3}}{\sigma C}\left(2b_{1,0}^{\left(2\right)}+b_{1,0}^{\prime\left(2\right)}\right),\label{eq:B00Y}
\end{equation}

\begin{equation}
\mathcal{B}_{2,0}^{Y}=\frac{2\sqrt{15}}{\sigma C}\left(-b_{1,0}^{\left(2\right)}-\sqrt{21}b_{3,0}^{\left(2\right)}+b_{1,0}^{\prime\left(2\right)}\right),\label{eq:B20Y}
\end{equation}

\begin{equation}
\mathcal{B}_{2,0}^{\Psi}=\frac{\sqrt{30}}{\sigma C}\left(-\sqrt{3}b_{1,0}^{\left(2\right)}+2\sqrt{7}b_{3,0}^{\left(2\right)}+\sqrt{3}b_{1,0}^{\prime\left(2\right)}\right).\label{eq:B20psi}
\end{equation}

These coefficients quantify the contributions of the fields $\vec{Y}_{0,0}^{\mathrm{M}}(\hat{k}^{\mathrm{M}})$$\propto$$\hat{k}^{\mathrm{M}}$,
$\vec{Y}_{2,0}^{\mathrm{M}}(\hat{k}^{\mathrm{M}})$$\propto$$(3\cos^{2}\theta_{k}^{\mathrm{M}}$$-1)\hat{k}^{\mathrm{M}}$,
and $\vec{\Psi}_{2,0}^{\mathrm{M}}(\hat{k}^{\mathrm{M}})$$\propto$$-\cos\theta_{k}^{\mathrm{M}}$$\sin\theta_{k}^{\mathrm{M}}\hat{\theta}_{k}^{\mathrm{M}}$
to the total field $\vec{B}^{\mathrm{M}}(\hat{k}^{\mathrm{M}})$ {[}Eq.
(\ref{eq:B}){]}. Equations (\ref{eq:B00Y})-(\ref{eq:B20psi}) thus
clearly show how structural information of the molecular field $\vec{B}^{\mathrm{M}}(\vec{k}^{\mathrm{M}})$
can be reconstructed from photoelectron angular distributions resulting
from an initially isotropic sample of chiral molecules.

\section{Conclusions\label{sec:Conclusions}}

We have presented an alternative approach to obtain expressions for
the $b_{l,m}^{\left(N\right)}$ coefficients of photoelectron angular
distributions resulting from perturbative $N$-photon ionization of
isotropic samples. These expressions are explicitly written in terms
of products between the molecular rotational invariants and the setup
rotational invariants, and do not invoke a partial wave expansion
for the scattering wave function. The molecular rotational invariants
are expressed in terms of vector products involving only molecular
vectors: transition dipoles and the photoelectron momentum labeling
a particular scattering state in the molecular frame. The setup rotational
invariants are expressed in terms of vector products involving only
setup vectors: field polarization vectors and detection axes. Our
expressions reveal the coupling of molecular and setup rotational
invariants. Knowledge of this coupling can assist the interpretation
and design of future experiments and simulations. The standard expressions
can be recovered by subsequent expansion of the scattering wave function
if needed. 

With the help of this methodology we found that, independently of
the polarization of the field, enantiosensitive $b_{l,m}^{\left(N_{1},N_{2}\right)}$
coefficients resulting from interference between pathways involving
$N_{1}$ and $N_{2}$ photons have odd $l+N_{1}+N_{2}$. 

The application of our methodology to the case of one-photon ionization
$\vert0\rangle$$\rightarrow$$\vert\vec{k}\rangle$ reveals a clear
meaning for the molecular information encoded in each of the $b_{l,m}^{\left(1\right)}$
coefficients which is otherwise obscured in the usual (and equivalent)
formulation in terms of partial waves: $b_{0,0}^{\left(1\right)}$
encodes the average magnitude of the photoionization dipole $\vec{D}(\vec{k})\equiv\langle\vec{k}\vert\vec{d}\vert0\rangle$;
$b_{1,0}^{\left(1\right)}$ encodes the average radial component of
the propensity field $\vec{B}\equiv i\vec{D}^{*}\times\vec{D},$ which
in turn encodes the average circular dichroism of $\vec{D}(\vec{k})$
and depends only on its transverse components; and $b_{2,0}^{\left(1\right)}$
encodes the average radial component of $\vec{D}(\vec{k})$. The averages
are taken with respect to the direction of the photoelectron momentum
$\vec{k}$ in the molecular frame.  $b_{1,0}^{\left(1\right)}$ is
sensitive to a single coefficient of the vector spherical harmonic
expansion of $\vec{B}(\vec{k})$.

We also derived expressions for the coefficients $b_{0,0}^{\left(2\right)}$,
$b_{1,0}^{\left(2\right)}$, and $b_{3,0}^{\left(2\right)}$ relevant
for two-photon resonantly enhanced ionization $\vert0\rangle$$\rightarrow$$\vert1\rangle$$\rightarrow$$\vert\vec{k}\rangle$
of isotropic chiral samples with circularly polarized light. The coefficients
$b_{0,0}^{\left(2\right)}$ and $b_{1,0}^{\left(2\right)}$ have analogous
interpretations to those found in the one-photon case provided one
takes into account an effective anisotropic stretching of the photoionization
dipoles. $b_{1,0}^{\left(2\right)}$ and $b_{3,0}^{\left(2\right)}$
yield structural information about the propensity field $\vec{B}\equiv i\vec{D}^{*}\times\vec{D}$
\cite{ordonez_propensity_2019}, which encodes the $\vec{k}$-specific
circular dichroism. In particular they depend only on three coefficients
of the vector spherical harmonic expansion of $\vec{B}(\vec{k})$.
These coefficients can be solved for in terms of $b_{1,0}^{\left(2\right)}$,
$b_{3,0}^{\left(2\right)}$, and $b_{1,0}^{\prime\left(2\right)}$,
where the latter corresponds to the process where the first photon
is linearly polarized. 

Further application of the methodology introduced here can be found
in the companion paper \cite{ordonez_tensorial}, where it is used
to analyze the enantiosensitive asymmetry recently found in the photoelectron
angular distributions resulting from interaction of chiral samples
with a field containing $\omega$ and $2\omega$ frequencies linearly
polarized orthogonal to each other \cite{demekhin_photoelectron_2018,demekhin_photoelectron_2019}.

\section{Appendix}

\subsection{Real spherical harmonics}

The real spherical harmonics (with tilde) are defined in terms of
the complex spherical harmonics (without tilde) according to

\begin{equation}
\tilde{Y}_{l}^{m}=\begin{cases}
\sqrt{2}\left(-1\right)^{m}\mathrm{Im}\left\{ Y_{l}^{\left|m\right|}\right\} , & m<0,\\
Y_{l}^{0} & m=0,\\
\sqrt{2}\left(-1\right)^{m}\mathrm{Re}\left\{ Y_{l}^{\left|m\right|}\right\} , & m>0,
\end{cases}
\end{equation}

and satisfy the orthonormality relation

\begin{equation}
\int\mathrm{d}\Omega\,\tilde{Y}_{l}^{m}\tilde{Y}_{\lambda}^{\mu}=\delta_{l,\lambda}\delta_{m,\mu}.
\end{equation}

For an arbitrary function $W$, the relation between the coefficients
of the real and the complex spherical harmonics can be derived from

\begin{equation}
W=\sum_{l,m}b_{l,m}Y_{l}^{m}=\sum_{l}\left\{ \tilde{b}_{l,0}\tilde{Y}_{l}^{0}+\sum_{\left|m\right|=1}^{l}\left[\tilde{b}_{l,\left|m\right|}Y_{l}^{\left|m\right|}+b_{l,-\left|m\right|}Y_{l}^{-\left|m\right|}\right]\right\} ,
\end{equation}

and yields

\begin{equation}
\tilde{b}_{l,m}=\begin{cases}
-\left(-1\right)^{m}\sqrt{2}\mathrm{Im}\left\{ b_{l,\left|m\right|}\right\} , & m<0,\\
b_{l,m}, & m=0,\\
\left(-1\right)^{m}\sqrt{2}\mathrm{Re}\left\{ b_{l,m}\right\} , & m>0.
\end{cases}
\end{equation}

\subsection{Derivation of the $\boldsymbol{\tilde{b}_{l,m}}$ coefficients in
one-photon-ionization}

According to Eqs. \eqref{eq:b_lm_puzzling}, \eqref{eq:field}, and
\eqref{eq:a1}, and following Ref. \cite{andrews_threedimensional_1977}
for the orientation averaging, we obtain

\begin{align}
b_{0,0}^{\left(1\right)} & =\left|A^{\left(1\right)}\right|^{2}\int\mathrm{d}\Omega_{k}^{\mathrm{M}}\int\mathrm{d}\varrho\,Y_{0}^{0}(\hat{k}^{\mathrm{L}})\left(\vec{D}^{\mathrm{L}*}\cdot\vec{F}^{\mathrm{L}*}\right)\left(\vec{D}^{\mathrm{L}}\cdot\vec{F}^{\mathrm{L}}\right),\nonumber \\
 & =\frac{\left|A^{\left(1\right)}\right|^{2}}{\sqrt{4\pi}}\int\mathrm{d}\Omega_{k}^{\mathrm{M}}\int\mathrm{d}\varrho\,\left(\vec{D}^{\mathrm{L}*}\cdot\vec{F}^{\mathrm{L}*}\right)\left(\vec{D}^{\mathrm{L}}\cdot\vec{F}^{\mathrm{L}}\right),\nonumber \\
 & =\frac{\left|A^{\left(1\right)}\right|^{2}}{3\sqrt{4\pi}}\int\mathrm{d}\Omega_{k}^{\mathrm{M}}\left|\vec{D}^{\mathrm{M}}\right|^{2}\left|\vec{F}^{\mathrm{L}}\right|^{2}.\label{eq:b_00_appendix}
\end{align}

\begin{align}
b_{1,0}^{\left(1\right)} & =\left|A^{\left(1\right)}\right|^{2}\int\mathrm{d}\Omega_{k}^{\mathrm{M}}\int\mathrm{d}\varrho\,Y_{1}^{0}(\hat{k}^{\mathrm{L}})\left(\vec{D}^{\mathrm{L}*}\cdot\vec{F}^{\mathrm{L}*}\right)\left(\vec{D}^{\mathrm{L}}\cdot\vec{F}^{\mathrm{L}}\right),\nonumber \\
 & =\left|A^{\left(1\right)}\right|^{2}\sqrt{\frac{3}{4\pi}}\int\mathrm{d}\Omega_{k}^{\mathrm{M}}\int\mathrm{d}\varrho\,\left(\hat{k}^{\mathrm{L}}\cdot\hat{z}^{\mathrm{L}}\right)\left(\vec{D}^{\mathrm{L}*}\cdot\vec{F}^{\mathrm{L}*}\right)\left(\vec{D}^{\mathrm{L}}\cdot\vec{F}^{\mathrm{L}}\right),\nonumber \\
 & =\left|A^{\left(1\right)}\right|^{2}\frac{1}{6}\sqrt{\frac{3}{4\pi}}\int\mathrm{d}\Omega_{k}^{\mathrm{M}}\left[\hat{k}^{\mathrm{M}}\cdot\left(\vec{D}^{\mathrm{M}*}\times\vec{D}^{\mathrm{M}}\right)\right]\left[\hat{z}^{\mathrm{L}}\cdot\left(\vec{F}^{\mathrm{L}*}\times\vec{F}^{\mathrm{L}}\right)\right].\label{eq:b10_appendix}
\end{align}

\begin{align}
b_{2,0}^{\left(1\right)} & =\left|A^{\left(1\right)}\right|^{2}\int\mathrm{d}\Omega_{k}^{\mathrm{M}}\int\mathrm{d}\varrho\,Y_{2}^{0}(\hat{k}^{\mathrm{L}})\left(\vec{D}^{\mathrm{L}*}\cdot\vec{F}^{\mathrm{L}*}\right)\left(\vec{D}^{\mathrm{L}}\cdot\vec{F}^{\mathrm{L}}\right),\nonumber \\
 & =\left|A^{\left(1\right)}\right|^{2}\frac{1}{4}\sqrt{\frac{5}{\pi}}\int\mathrm{d}\Omega_{k}^{\mathrm{M}}\int\mathrm{d}\varrho\,\left[3\left(\hat{k}^{\mathrm{L}}\cdot\hat{z}^{\mathrm{L}}\right)^{2}-1\right]\left(\vec{D}^{\mathrm{L}*}\cdot\vec{F}^{\mathrm{L}*}\right)\left(\vec{D}^{\mathrm{L}}\cdot\vec{F}^{\mathrm{L}}\right)\nonumber \\
 & =\left|A^{\left(1\right)}\right|^{2}\frac{3}{4}\sqrt{\frac{5}{\pi}}\int\mathrm{d}\Omega_{k}^{\mathrm{M}}\int\mathrm{d}\varrho\,\left(\hat{k}^{\mathrm{L}}\cdot\hat{z}^{\mathrm{L}}\right)^{2}\left(\vec{D}^{\mathrm{L}*}\cdot\vec{F}^{\mathrm{L}*}\right)\left(\vec{D}^{\mathrm{L}}\cdot\vec{F}^{\mathrm{L}}\right)-\sqrt{\frac{5}{4}}b_{0,0}^{\left(1\right)},\label{eq:b20_}
\end{align}

For the remaining integral over orientations in $b_{2,0}^{\left(1\right)}$
we have

\begin{align}
\int\mathrm{d}\varrho\,\left(\hat{k}^{\mathrm{L}}\cdot\hat{z}^{\mathrm{L}}\right)^{2}\left(\vec{D}^{\mathrm{L}*}\cdot\vec{F}^{\mathrm{L}*}\right)\left(\vec{D}^{\mathrm{L}}\cdot\vec{F}^{\mathrm{L}}\right) & =\vec{g}^{\left(4\right)}\cdot M^{\left(4\right)}\vec{f}^{\left(4\right)}\label{eq:b20_integral}
\end{align}

where

\textbf{
\begin{equation}
\vec{g}^{\left(4\right)}=\left[\begin{array}{c}
(\hat{k}^{\mathrm{M}}\cdot\vec{k}^{\mathrm{M}})(\vec{D}^{\mathrm{M}*}\cdot\vec{D}^{\mathrm{M}})\\
(\hat{k}^{\mathrm{M}}\cdot\vec{D}^{\mathrm{M}*})(\hat{k}^{\mathrm{M}}\cdot\vec{D}^{\mathrm{M}})\\
(\hat{k}^{\mathrm{M}}\cdot\vec{D}^{\mathrm{M}})(\hat{k}^{\mathrm{M}}\cdot\vec{D}^{\mathrm{M}*})
\end{array}\right]=\left[\begin{array}{c}
\vert\vec{D}^{\mathrm{M}}\vert^{2}\\
\vert\hat{k}^{\mathrm{M}}\cdot\vec{D}^{\mathrm{M}}\vert^{2}\\
\vert\hat{k}^{\mathrm{M}}\cdot\vec{D}^{\mathrm{M}}\vert^{2}
\end{array}\right]\label{eq:g4}
\end{equation}
}

\begin{equation}
M^{\left(4\right)}=\frac{1}{30}\left[\begin{array}{ccc}
4 & -1 & -1\\
-1 & 4 & -1\\
-1 & -1 & 4
\end{array}\right]\label{eq:M4}
\end{equation}

\begin{equation}
\vec{f}^{\left(4\right)}=\left[\begin{array}{c}
\left(\hat{z}^{\mathrm{L}}\cdot\hat{z}^{\mathrm{L}}\right)(\vec{F}^{\mathrm{L}*}\cdot\vec{F}^{\mathrm{L}})\\
(\hat{z}^{\mathrm{L}}\cdot\vec{F}^{\mathrm{L}*})(\hat{z}^{\mathrm{L}}\cdot\vec{F}^{\mathrm{L}})\\
(\hat{z}^{\mathrm{L}}\cdot\vec{F}^{\mathrm{L}})(\hat{z}^{\mathrm{L}}\cdot\vec{F}^{\mathrm{L}*})
\end{array}\right]=\left[\begin{array}{c}
\vert\vec{F}^{\mathrm{L}}\vert^{2}\\
\vert\hat{z}^{\mathrm{L}}\cdot\vec{F}^{\mathrm{L}}\vert^{2}\\
\vert\hat{z}^{\mathrm{L}}\cdot\vec{F}^{\mathrm{L}}\vert^{2}
\end{array}\right]\label{eq:f4}
\end{equation}

Replacing Eqs. \eqref{eq:g4}, \eqref{eq:M4}, \eqref{eq:f4} in Eq.
\eqref{eq:b20_integral} we get

\begin{multline}
\int\mathrm{d}\varrho\,\left(\hat{k}^{\mathrm{L}}\cdot\hat{z}^{\mathrm{L}}\right)^{2}\left(\vec{D}^{\mathrm{L}*}\cdot\vec{F}^{\mathrm{L}*}\right)\left(\vec{D}^{\mathrm{L}}\cdot\vec{F}^{\mathrm{L}}\right)\\
=\frac{1}{15}\left\{ \left[2\left|\vec{D}^{\mathrm{M}}\right|^{2}-\left|\hat{k}^{\mathrm{M}}\cdot\vec{D}^{\mathrm{M}}\right|^{2}\right]\left|\vec{F}^{\mathrm{L}}\right|^{2}-\left[\left|\vec{D}^{\mathrm{M}}\right|^{2}-3\left|\hat{k}^{\mathrm{M}}\cdot\vec{D}^{\mathrm{M}}\right|^{2}\right]\left|\hat{z}^{\mathrm{L}}\cdot\vec{F}^{\mathrm{L}}\right|^{2}\right\} ,\label{eq:b20_integral_2}
\end{multline}

and replacing Eq. \eqref{eq:b20_integral_2} in Eq. \eqref{eq:b20_}
we arrive to the rather symmetric result

\begin{align}
b_{2,0}^{\left(1\right)} & =\frac{\left|A^{\left(1\right)}\right|^{2}}{12\sqrt{5\pi}}\int\mathrm{d}\Omega_{k}^{\mathrm{M}}\left(3\left|\hat{k}^{\mathrm{M}}\cdot\vec{D}^{\mathrm{M}}\right|^{2}-\left|\vec{D}^{\mathrm{M}}\right|^{2}\right)\left(3\left|\hat{z}^{\mathrm{L}}\cdot\vec{F}^{\mathrm{L}}\right|^{2}-\left|\vec{F}^{\mathrm{L}}\right|^{2}\right)\label{eq:b20_appendix}
\end{align}

Similarly, by replacing $\hat{z}^{\mathrm{L}}$ by either $\hat{x}^{\mathrm{L}}$
or $\hat{y}^{\mathrm{L}}$ in Eq. \eqref{eq:b20_integral_2} we obtain

\begin{align}
\tilde{b}_{2,2}^{\left(1\right)} & =\left|A^{\left(1\right)}\right|^{2}\int\mathrm{d}\Omega_{k}^{\mathrm{M}}\int\mathrm{d}\varrho\,\tilde{Y}_{2}^{2}(\hat{k}^{\mathrm{L}})\left(\vec{D}^{\mathrm{L}*}\cdot\vec{F}^{\mathrm{L}*}\right)\left(\vec{D}^{\mathrm{L}}\cdot\vec{F}^{\mathrm{L}}\right),\nonumber \\
 & =\left|A^{\left(1\right)}\right|^{2}\frac{1}{4}\sqrt{\frac{15}{\pi}}\int\mathrm{d}\Omega_{k}^{\mathrm{M}}\int\mathrm{d}\varrho\,\left[\left(\hat{k}^{\mathrm{L}}\cdot\hat{x}^{\mathrm{L}}\right)^{2}-\left(\hat{k}^{\mathrm{L}}\cdot\hat{y}^{\mathrm{L}}\right)^{2}\right]\left(\vec{D}^{\mathrm{L}*}\cdot\vec{F}^{\mathrm{L}*}\right)\left(\vec{D}^{\mathrm{L}}\cdot\vec{F}^{\mathrm{L}}\right)\nonumber \\
 & =\left|A^{\left(1\right)}\right|^{2}\frac{1}{4\sqrt{15\pi}}\int\mathrm{d}\Omega_{k}^{\mathrm{M}}\left[3\left|\hat{k}^{\mathrm{M}}\cdot\vec{D}^{\mathrm{M}}\right|^{2}-\left|\vec{D}^{\mathrm{M}}\right|^{2}\right]\left[\left|\hat{x}^{\mathrm{L}}\cdot\vec{F}^{\mathrm{L}}\right|^{2}-\left|\hat{y}^{\mathrm{L}}\cdot\vec{F}^{\mathrm{L}}\right|^{2}\right].
\end{align}

Finally, 

\begin{align}
\tilde{b}_{2,-2}^{\left(1\right)}\left(k\right) & =\left|A^{\left(1\right)}\right|^{2}\int\mathrm{d}\Omega_{k}^{\mathrm{M}}\int\mathrm{d}\varrho\,\tilde{Y}_{2}^{-2}(\hat{k}^{\mathrm{L}})\left(\vec{D}^{\mathrm{L}*}\cdot\vec{F}^{\mathrm{L}*}\right)\left(\vec{D}^{\mathrm{L}}\cdot\vec{F}^{\mathrm{L}}\right),\nonumber \\
 & =\left|A^{\left(1\right)}\right|^{2}\frac{1}{2}\sqrt{\frac{15}{\pi}}\int\mathrm{d}\Omega_{k}^{\mathrm{M}}\int\mathrm{d}\varrho\,\left(\hat{k}^{\mathrm{L}}\cdot\hat{x}^{\mathrm{L}}\right)\left(\hat{k}^{\mathrm{L}}\cdot\hat{y}^{\mathrm{L}}\right)\left(\vec{D}^{\mathrm{L}*}\cdot\vec{F}^{\mathrm{L}*}\right)\left(\vec{D}^{\mathrm{L}}\cdot\vec{F}^{\mathrm{L}}\right)\nonumber \\
 & =\left|A^{\left(1\right)}\right|^{2}\frac{1}{2}\sqrt{\frac{15}{\pi}}\int\mathrm{d}\Omega_{k}^{\mathrm{M}}\frac{1}{30}\left[3\vert\hat{k}^{\mathrm{M}}\cdot\vec{D}^{\mathrm{M}}\vert^{2}-\vert\vec{D}^{\mathrm{M}}\vert^{2}\right]\left[(\hat{x}^{\mathrm{L}}\cdot\vec{F}^{\mathrm{L}*})(\hat{y}^{\mathrm{L}}\cdot\vec{F}^{\mathrm{L}})+\mathrm{c.c.}\right]\nonumber \\
 & =0
\end{align}

\subsection{Range of values of $\boldsymbol{b_{1,0}}$ in one-photon PECD}

For circularly polarized light we have $\vec{F}=F_{+}\hat{\epsilon}_{+}$.
The $b_{l,0}^{\left(1\right)}$ coefficients take the form ($\tilde{b}_{2,2}^{\left(1\right)}=0$)

\begin{equation}
b_{0,0}^{\left(1\right)}=\left|A^{\left(1\right)}\right|^{2}\left\{ \frac{1}{3\sqrt{4\pi}}\int\mathrm{d}\Omega_{k}^{\mathrm{M}}\left(\left|D_{+}^{\mathrm{M}}\right|^{2}+\left|D_{-}^{\mathrm{M}}\right|^{2}+\left|D_{k}^{\mathrm{M}}\right|^{2}\right)\right\} \left|F\right|^{2},
\end{equation}

\begin{equation}
b_{1,0}^{\left(1\right)}=\left|A^{\left(1\right)}\right|^{2}\left\{ \frac{1}{6}\sqrt{\frac{3}{4\pi}}\int\mathrm{d}\Omega_{k}^{\mathrm{M}}\left(\left|D_{+}^{\mathrm{M}}\right|^{2}-\left|D_{-}^{\mathrm{M}}\right|^{2}\right)\right\} \left|F\right|^{2},
\end{equation}

\begin{equation}
b_{2,0}^{\left(1\right)}=\left|A^{\left(1\right)}\right|^{2}\left\{ \frac{1}{12\sqrt{5\pi}}\int\mathrm{d}\Omega_{k}^{\mathrm{M}}\left(\left|D_{+}^{\mathrm{M}}\right|^{2}+\left|D_{-}^{\mathrm{M}}\right|^{2}-2\left|D_{k}^{\mathrm{M}}\right|^{2}\right)\right\} \left|F\right|^{2}.
\end{equation}

The sum and the difference of the $\vec{k}$-averaged absolute value
squares of the spherical components of $\vec{D}$ can be written in
terms of $b_{0,0}^{\left(1\right)}$, $b_{1,0}^{\left(2\right)}$,
and $b_{2,0}^{\left(1\right)}$ as

\begin{equation}
\int\mathrm{d}\Omega_{k}^{\mathrm{M}}\left(\left|D_{+}^{\mathrm{M}}\right|^{2}+\left|D_{-}^{\mathrm{M}}\right|^{2}\right)=\frac{2\left(\sqrt{4\pi}b_{0,0}^{\left(1\right)}+2\sqrt{5\pi}b_{2,0}^{\left(1\right)}\right)}{\left|A^{\left(1\right)}\right|^{2}\left|F\right|^{2}}
\end{equation}

\begin{equation}
\int\mathrm{d}\Omega_{k}^{\mathrm{M}}\left(\left|D_{+}^{\mathrm{M}}\right|^{2}-\left|D_{-}^{\mathrm{M}}\right|^{2}\right)=\frac{6\sqrt{\frac{4\pi}{3}}b_{1,0}^{\left(1\right)}}{\left|A^{\left(1\right)}\right|^{2}\left|F\right|^{2}}
\end{equation}

Since 

\begin{align}
\left|\int\mathrm{d}\Omega_{k}^{\mathrm{M}}\left(\left|D_{+}^{\mathrm{M}}\right|^{2}-\left|D_{-}^{\mathrm{M}}\right|^{2}\right)\right| & \leq\int\mathrm{d}\Omega_{k}^{\mathrm{M}}\left|\left|D_{+}^{\mathrm{M}}\right|^{2}-\left|D_{-}^{\mathrm{M}}\right|^{2}\right|,\nonumber \\
 & \leq\int\mathrm{d}\Omega_{k}^{\mathrm{M}}\left(\left|D_{+}^{\mathrm{M}}\right|^{2}+\left|D_{-}^{\mathrm{M}}\right|^{2}\right),
\end{align}

then 

\begin{equation}
\sqrt{3}\left|b_{1,0}^{\left(1\right)}\right|\leq\left(b_{0,0}^{\left(1\right)}+\sqrt{5}b_{2,0}^{\left(1\right)}\right).
\end{equation}

\subsection{Derivation of the $\boldsymbol{b_{0,0}}$, $\boldsymbol{b_{1,0}}$,
and $\boldsymbol{b_{3,0}}$ coefficients in two-photon PECD}

From Eq. \eqref{eq:b_lm_tensor} we have that

\begin{equation}
b_{0,0}^{\left(2\right)}=\frac{1}{\sqrt{4\pi}}\left|A^{\left(2\right)}\right|^{2}\int\mathrm{d}\Omega_{k}^{\mathrm{M}}\int\mathrm{d}\varrho\,\left|\vec{D}^{\mathrm{L}}\cdot\vec{F}^{\mathrm{L}}\right|^{2}\left|\vec{d}^{\mathrm{L}}\cdot\vec{F}^{\mathrm{L}}\right|^{2},\label{eq:b00_2photon_appendix}
\end{equation}

where we use the shorthand notation $\vec{D}^{\mathrm{L}}\equiv\vec{d}_{\vec{k}^{\mathrm{M}},1}^{\mathrm{L}}$,
$\vec{d}^{\mathrm{L}}\equiv\vec{d}_{1,0}^{\mathrm{L}}$, and $\vec{F}^{\mathrm{L}}\equiv\vec{F}_{\omega_{L}}^{\mathrm{L}}$.
The orientation averaging can be performed following Ref. \cite{andrews_threedimensional_1977},

\begin{equation}
\int\mathrm{d}\varrho\,\left(\vec{D}^{\mathrm{L}}\cdot\vec{F}^{\mathrm{L}}\right)^{*}\left(\vec{D}^{\mathrm{L}}\cdot\vec{F}^{\mathrm{L}}\right)\left(\vec{d}^{\mathrm{L}}\cdot\vec{F}^{\mathrm{L}}\right)^{*}\left(\vec{d}^{\mathrm{L}}\cdot\vec{F}^{\mathrm{L}}\right)=\vec{g}^{\left(4\right)}\cdot M^{\left(4\right)}\vec{f}^{\left(4\right)},\label{eq:b00_2photon_integral}
\end{equation}

where\footnote{In the absence of magnetic fields $\vec{d}^{\mathrm{M}}$ can be taken
real.}

\textbf{
\begin{equation}
\vec{g}^{\left(4\right)}=\left[\begin{array}{c}
\vert\vec{D}^{\mathrm{M}}\vert^{2}d^{2}\\
\vert\vec{D}^{\mathrm{M}}\cdot\vec{d}^{\mathrm{M}}\vert^{2}\\
\vert\vec{D}^{\mathrm{M}}\cdot\vec{d}^{\mathrm{M}}\vert^{2}
\end{array}\right],\label{eq:g4_2photon}
\end{equation}
}

\begin{equation}
\vec{f}^{\left(4\right)}=\left[\begin{array}{c}
\vert\vec{F}^{\mathrm{L}}\vert^{4}\\
\vert(\vec{F}^{\mathrm{L}})^{2}\vert^{2}\\
\vert\vec{F}^{\mathrm{L}}\vert^{4}
\end{array}\right],\label{eq:f4_2photon}
\end{equation}

$M^{\left(4\right)}$ is given by Eq. \eqref{eq:M4}. Replacing Eqs.
\eqref{eq:M4}, \eqref{eq:b00_2photon_integral}, \eqref{eq:g4_2photon},
and \eqref{eq:f4_2photon} in Eq. \eqref{eq:b00_2photon_appendix}
yields

\begin{align}
b_{0,0}^{\left(2\right)} & =\frac{1}{\sqrt{4\pi}}\left|A^{\left(2\right)}\right|^{2}\frac{1}{30}\int\mathrm{d}\Omega_{k}^{\mathrm{M}}\bigg\{\left[\vert\vec{D}^{\mathrm{M}}\cdot\vec{d}^{\mathrm{M}}\vert^{2}+3\vert\vec{D}^{\mathrm{M}}\vert^{2}d^{2}\right]\vert\vec{F}^{\mathrm{L}}\vert^{4}.\nonumber \\
 & +\left[3\vert\vec{D}^{\mathrm{M}}\cdot\vec{d}^{\mathrm{M}}\vert^{2}-\vert\vec{D}^{\mathrm{M}}\vert^{2}d^{2}\right]\vert(\vec{F}^{\mathrm{L}})^{2}\vert^{2}\bigg\}\label{eq:b00_2photon_general}
\end{align}

This expression is valid for arbitrary $\vec{d}^{\mathrm{M}}$ and
arbitrary polarization. If we choose the molecular frame so that $\vec{d}^{\mathrm{M}}=d\hat{z}^{\mathrm{M}}$,
we focus on the case of circular polarization $\vec{F}^{\mathrm{L}}=F\left(\hat{x}^{\mathrm{L}}\pm i\hat{y}^{\mathrm{L}}\right)/\sqrt{2}$,
and use the definition (\ref{eq:Deff00-1}), Eq. \eqref{eq:b00_2photon_general}
reduces to Eq. \eqref{eq:b00_2}.

Similarly, for the case of $b_{1,0}^{\left(2\right)}$ we get {[}see
Eq. \eqref{eq:b_lm_tensor}{]}

\begin{align}
b_{1,0}^{\left(2\right)} & =\sqrt{\frac{3}{4\pi}}\left|A^{\left(2\right)}\right|^{2}\int\mathrm{d}\Omega_{k}^{\mathrm{M}}\int\mathrm{d}\varrho\,\left(\hat{k}^{\mathrm{L}}\cdot\hat{z}^{\mathrm{L}}\right)\left|\vec{D}^{\mathrm{L}}\cdot\vec{F}^{\mathrm{L}}\right|^{2}\left|\vec{d}^{\mathrm{L}}\cdot\vec{F}^{\mathrm{L}}\right|^{2}.\label{eq:b10(2)_appendix}
\end{align}

The integral over orientations $\varrho$ reads as

\begin{equation}
\int\mathrm{d}\varrho\,\left(\hat{k}^{\mathrm{L}}\cdot\hat{z}^{\mathrm{L}}\right)\left(\vec{D}^{\mathrm{L}}\cdot\vec{F}^{\mathrm{L}}\right)^{*}\left(\vec{d}^{\mathrm{L}}\cdot\vec{F}^{\mathrm{L}}\right)^{*}\left(\vec{D}^{\mathrm{L}}\cdot\vec{F}^{\mathrm{L}}\right)\left(\vec{d}^{\mathrm{L}}\cdot\vec{F}^{\mathrm{L}}\right)=\vec{g}^{\left(5\right)}\cdot M^{\left(5\right)}\vec{f}^{\left(5\right)},\label{eq:b10_orientation_averaging}
\end{equation}

where\footnote{For the moment we omit the $\mathrm{M}$ superscript on $\vec{k}$,
$\vec{D}$, $\vec{D}^{*}$, and $\vec{d}$; and the superscript $\mathrm{L}$
on $\hat{z}$, $\vec{F}$, and $\vec{F}^{*}$.}

\begin{equation}
\vec{g}^{\left(5\right)}=\left(\begin{array}{c}
\left[\hat{k}\cdot\left(\vec{D}^{*}\x\vec{d}\right)\right]\left(\vec{D}\t\vec{d}\right)\\
\left[\hat{k}\cdot\left(\vec{D}^{*}\x\vec{D}\right)\right]d^{2}\\
\left[\hat{k}\cdot\left(\vec{D}^{*}\x\vec{d}\right)\right]\left(\vec{D}\cdot\vec{d}\right)\\
\left[\hat{k}\cdot\left(\vec{d}\x\vec{D}\right)\right]\left(\vec{D}^{*}\t\vec{d}\right)\\
0\\
\left[\hat{k}\cdot\left(\vec{D}\x\vec{d}\right)\right]\left(\vec{D}^{*}\t\vec{d}\right)
\end{array}\right),\qquad\vec{f}^{\left(5\right)}=\left[\hat{z}\cdot\left(\vec{F}^{*}\x\vec{F}\right)\right]\left|\vec{F}\right|^{2}\left(\begin{array}{c}
0\\
1\\
1\\
1\\
1\\
0
\end{array}\right),
\end{equation}

\begin{equation}
M^{\left(5\right)}=\frac{1}{30}\left(\begin{array}{cccccc}
3 & -1 & -1 & 1 & 1 & 0\\
-1 & 3 & -1 & -1 & 0 & 1\\
-1 & -1 & 3 & 0 & -1 & -1\\
1 & -1 & 0 & 3 & -1 & 1\\
1 & 0 & -1 & -1 & 3 & -1\\
0 & 1 & -1 & 1 & -1 & 3
\end{array}\right).
\end{equation}

Since $M^{\left(5\right)}\vec{f}^{\left(5\right)}=\vec{f}^{\left(5\right)}$,
then 

\begin{align}
\vec{g}^{\left(5\right)}\cdot M^{\left(5\right)}\vec{f}^{\left(5\right)} & =\frac{1}{30}\left\{ \left[\hat{k}\cdot\left(\vec{D}^{*}\x\vec{D}\right)\right]d^{2}+\left[\hat{k}\cdot\left(\vec{D}^{*}\x\vec{d}\right)\right]\left(\vec{D}\cdot\vec{d}\right)+\left[\hat{k}\cdot\left(\vec{d}\x\vec{D}\right)\right]\left(\vec{D}^{*}\t\vec{d}\right)\right\} \nonumber \\
 & \times\left\{ \left[\hat{z}\cdot\left(\vec{F}^{*}\x\vec{F}\right)\right]\left|\vec{F}\right|^{2}\right\} \label{eq:gMf5}
\end{align}

With the help of some vector algebra the second and third terms can
be rewritten as 

\begin{equation}
\left[\hat{k}\t\left(\vec{D}^{*}\x\vec{d}\right)\right]\left(\vec{D}\cdot\vec{d}\right)-\left[\hat{k}\t\left(\vec{D}\x\vec{d}\right)\right]\left(\vec{D}^{*}\cdot\vec{d}\right)=d^{2}\left[\hat{k}-\left(\hat{k}\cdot\hat{d}\right)\hat{d}\right]\cdot\left(\vec{D}^{*}\x\vec{D}\right).\label{eq:nice_trick}
\end{equation}

Replacing Eqs. \eqref{eq:b10_orientation_averaging}-\eqref{eq:nice_trick}
in Eq. \eqref{eq:b10(2)_appendix} yields 

\begin{align}
b_{1,0}^{\left(2\right)} & =\sqrt{\frac{3}{4\pi}}\left|A^{\left(2\right)}\right|^{2}\frac{d^{2}\left|\vec{F}\right|^{2}}{30}\int\mathrm{d}\Omega_{k}^{\mathrm{M}}\left\{ \left[2\hat{k}-\left(\hat{k}\cdot\hat{d}\right)\hat{d}\right]\cdot\left(\vec{D}^{*}\x\vec{D}\right)\right\} \left\{ \left[\hat{z}\cdot\left(\vec{F}^{*}\x\vec{F}\right)\right]\right\} .\label{eq:b10_2photon_general}
\end{align}

This expression is valid for arbitrary orientations of $\vec{d}^{\mathrm{M}}$
and arbitrary polarization. If we choose the molecular frame so that
$\vec{d}^{\mathrm{M}}=d\hat{z}^{\mathrm{M}}$, focus on the case of
circular polarization $\vec{F}^{\mathrm{L}}=F\left(\hat{x}^{\mathrm{L}}\pm i\hat{y}^{\mathrm{L}}\right)/\sqrt{2}$,
and use definitions (\ref{eq:Deff10-1}) and (\ref{eq:K10}), Eq.
\eqref{eq:b10_2photon_general} reduces to Eqs. (\ref{eq:b10_2})
and (\ref{eq:b10_2_B})

Finally, for $b_{3,0}^{\left(2\right)}$ we get {[}see Eq. \eqref{eq:b_lm_tensor}{]}

\begin{align}
b_{3,0}^{\left(2\right)} & =\frac{5}{4}\sqrt{\frac{7}{\pi}}\left|A^{\left(2\right)}\right|^{2}\int\mathrm{d}\Omega_{k}^{\mathrm{M}}\int\mathrm{d}\varrho\,\left(\hat{k}^{\mathrm{L}}\cdot\hat{z}^{\mathrm{L}}\right)^{3}\left|\left(\vec{D}^{\mathrm{L}}\cdot\vec{F}^{\mathrm{L}}\right)\right|^{2}\left|\left(\vec{d}^{\mathrm{L}}\cdot\vec{F}^{\mathrm{L}}\right)\right|^{2}\nonumber \\
 & -\frac{3}{4}\sqrt{\frac{7}{\pi}}\sqrt{\frac{4\pi}{3}}b_{1,0}^{\left(2\right)}\label{eq:b30(2)}
\end{align}

The orientation integral in the first term reads as

\begin{equation}
\int\mathrm{d}\varrho\,\left(\hat{k}^{\mathrm{L}}\cdot\hat{z}^{\mathrm{L}}\right)^{3}\left(\vec{D}^{\mathrm{L}*}\cdot\vec{F}^{\mathrm{L}*}\right)\left(\vec{d}^{\mathrm{L}}\cdot\vec{F}^{\mathrm{L}*}\right)\left(\vec{D}^{\mathrm{L}}\cdot\vec{F}^{\mathrm{L}}\right)\left(\vec{d}^{\mathrm{L}}\cdot\vec{F}^{\mathrm{L}}\right)=\vec{g}^{\left(7\right)}\cdot M^{\left(7\right)}\vec{f}^{\left(7\right)}\label{eq:gMf7}
\end{equation}

From table III in Ref. \cite{andrews_threedimensional_1977} we see
that $f_{i}^{\left(7\right)}=g_{i}^{\left(7\right)}=0$ for $1\leq i\leq27$.
For $28\leq i\leq36$ we get\footnote{For the moment we omit the $\mathrm{M}$ superscript on $\vec{k}$,
$\vec{D}$, $\vec{D}^{*}$, and $\vec{d}$; and the superscript $\mathrm{L}$
on $\hat{z}$, $\vec{F}$, and $\vec{F}^{*}$.}

\begin{equation}
\vec{g}^{\left(7\right)}=\left[\begin{array}{c}
\left[\hat{k}\cdot\left(\vec{D}^{*}\times\vec{d}\right)\right]\left(\hat{k}\cdot\hat{k}\right)\left(\vec{D}\cdot\vec{d}\right)\\
\left[\hat{k}\cdot\left(\vec{D}^{*}\times\vec{d}\right)\right]\left(\hat{k}\cdot\vec{d}\right)\left(\hat{k}\cdot\vec{D}\right)\\
\left[\hat{k}\cdot\left(\vec{D}^{*}\times\vec{D}\right)\right]\left(\hat{k}\cdot\hat{k}\right)\left(\vec{d}\cdot\vec{d}\right)\\
\left[\hat{k}\cdot\left(\vec{D}^{*}\times\vec{D}\right)\right]\left(\hat{k}\cdot\vec{d}\right)\left(\hat{k}\cdot\vec{d}\right)\\
\left[\hat{k}\cdot\left(\vec{D}^{*}\times\vec{d}\right)\right]\left(\hat{k}\cdot\hat{k}\right)\left(\vec{d}\cdot\vec{D}\right)\\
\left[\hat{k}\cdot\left(\vec{D}^{*}\times\vec{d}\right)\right]\left(\hat{k}\cdot\vec{D}\right)\left(\hat{k}\cdot\vec{d}\right)\\
\left[\hat{k}\cdot\left(\vec{d}\times\vec{D}\right)\right]\left(\hat{k}\cdot\hat{k}\right)\left(\vec{D}^{*}\cdot\vec{d}\right)\\
0\\
\left[\hat{k}\cdot\left(\vec{D}\times\vec{d}\right)\right]\left(\hat{k}\cdot\hat{k}\right)\left(\vec{D}^{*}\cdot\vec{d}\right)
\end{array}\right]\equiv\left[\begin{array}{c}
g_{1}\\
g_{2}\\
g_{3}\\
g_{4}\\
g_{1}\\
g_{2}\\
-g_{1}^{*}\\
0\\
g_{1}^{*}
\end{array}\right]
\end{equation}

\begin{equation}
\vec{f}^{\left(7\right)}=\left[\begin{array}{c}
0\\
0\\
\left[\hat{z}\cdot\left(\vec{F}^{*}\times\vec{F}\right)\right]\left(\hat{z}\cdot\hat{z}\right)\left(\vec{F}^{*}\cdot\vec{F}\right)\\
\left[\hat{z}\cdot\left(\vec{F}^{*}\times\vec{F}\right)\right]\left(\hat{z}\cdot\vec{F}\right)\left(\hat{z}\cdot\vec{F}^{*}\right)\\
\left[\hat{z}\cdot\left(\vec{F}^{*}\times\vec{F}\right)\right]\left(\hat{z}\cdot\hat{z}\right)\left(\vec{F}^{*}\cdot\vec{F}\right)\\
\left[\hat{z}\cdot\left(\vec{F}^{*}\times\vec{F}\right)\right]\left(\hat{z}\cdot\vec{F}\right)\left(\hat{z}\cdot\vec{F}^{*}\right)\\
\left[\hat{z}\cdot\left(\vec{F}^{*}\times\vec{F}\right)\right]\left(\hat{z}\cdot\hat{z}\right)\left(\vec{F}^{*}\cdot\vec{F}\right)\\
\left[\hat{z}\cdot\left(\vec{F}^{*}\times\vec{F}\right)\right]\left(\hat{z}\cdot\hat{z}\right)\left(\vec{F}^{*}\cdot\vec{F}\right)\\
0
\end{array}\right]=\left[\hat{z}\cdot\left(\vec{F}^{*}\times\vec{F}\right)\right]\left[\begin{array}{c}
0\\
0\\
\left|\vec{F}\right|^{2}\\
\left|F_{z}\right|^{2}\\
\left|\vec{F}\right|^{2}\\
\left|F_{z}\right|^{2}\\
\left|\vec{F}\right|^{2}\\
\left|\vec{F}\right|^{2}\\
0
\end{array}\right]
\end{equation}

The relevant part of $M^{\left(7\right)}$ in Ref. \cite{andrews_threedimensional_1977}
reads as

\begin{equation}
M^{\left(7\right)}=\frac{1}{420}\left[\begin{array}{ccccccccc}
51 & -33 & -21 & 15 & -21 & 15 & 18 & 18 & 0\\
-33 & 45 & 15 & -15 & 15 & -15 & -12 & -12 & 0\\
-21 & 15 & 51 & -33 & -21 & 15 & -18 & 0 & 18\\
15 & -15 & -33 & 45 & 15 & -15 & 12 & 0 & -12\\
-21 & 15 & -21 & 15 & 51 & -33 & 0 & -18 & -18\\
15 & -15 & 15 & -15 & -33 & 45 & 0 & 12 & 12\\
18 & -12 & -18 & 12 & 0 & 0 & 30 & -6 & 6\\
18 & -12 & 0 & 0 & -18 & 12 & -6 & 30 & -6\\
0 & 0 & 18 & -12 & -18 & 12 & 6 & -6 & 30
\end{array}\right],
\end{equation}

therefore

\begin{equation}
\vec{g}^{\left(7\right)}\cdot M^{\left(7\right)}\vec{f}^{\left(7\right)}=\frac{1}{70}\left[\left(2i\mathrm{Im}\left\{ g_{1}\right\} +2g_{3}-g_{4}\right)\left|F\right|^{2}+\left(4i\mathrm{Im}\left\{ g_{1}\right\} -3g_{3}+5g_{4}\right)\left|F_{z}\right|^{2}\right]\left[\hat{z}\cdot\left(\vec{F}^{*}\times\vec{F}\right)\right].\label{eq:gMf7_solved}
\end{equation}

Equations \eqref{eq:b10_2photon_general}, \eqref{eq:b30(2)}, \eqref{eq:gMf7},
and \eqref{eq:gMf7_solved} yield 

\begin{align}
b_{3,0}^{\left(2\right)} & =\frac{1}{4}\sqrt{\frac{7}{\pi}}\left|A^{\left(2\right)}\right|^{2}\frac{1}{70}\int\mathrm{d}\Omega_{k}^{\mathrm{M}}d^{2}\left\{ \left[\left(1-5\left(\hat{k}\cdot\hat{d}\right)^{2}\right)\hat{k}+2\left(\hat{k}\cdot\hat{d}\right)\hat{d}\right]\cdot\left(\vec{D}^{*}\times\vec{D}\right)\right\} \nonumber \\
 & \times\left\{ \left[\hat{z}^{\mathrm{L}}\cdot\left(\vec{F}^{\mathrm{L}*}\times\vec{F}^{\mathrm{L}}\right)\right]\left(\left|\vec{F}^{\mathrm{L}}\right|^{2}-5\left|F_{z}^{\mathrm{L}}\right|^{2}\right)\right\} \label{eq:b30_2photon_general}
\end{align}

This expression is valid for arbitrary orientations of $\vec{d}^{\mathrm{M}}$
and arbitrary polarization. If we choose the molecular frame so that
$\vec{d}^{\mathrm{M}}=d\hat{z}^{\mathrm{M}}$, we focus on the case
of circular polarization $\vec{F}^{\mathrm{L}}=F\left(\hat{x}^{\mathrm{L}}\pm i\hat{y}^{\mathrm{L}}\right)/\sqrt{2}$,
and we use definition (\ref{eq:K30}), Eq. \eqref{eq:b30_2photon_general}
reduces to Eq. \eqref{eq:b30_2}.

\subsection{Derivation of $\boldsymbol{b_{1,0}^{\prime\left(2\right)}}$ in Eq.
(\ref{eq:b10_2_lin})}

Analogously to Eq. (\ref{eq:b_lm_2_circ}), the $b_{1,0}^{\prime\left(2\right)}$
coefficient corresponding to the process where the first photon is
linearly polarized along $\hat{z}^{\mathrm{L}}$ and the second photon
is circularly polarized in the $\hat{x}^{\mathrm{L}}\hat{y}^{\mathrm{L}}$
plane is given by 

\begin{equation}
b_{1,0}^{\prime\left(2\right)}\left(k\right)=\vert A^{\left(2\right)}\vert^{2}d^{2}\left|F\right|^{2}\int\mathrm{d}\Omega_{k}^{\mathrm{M}}\int\mathrm{d}\varrho\,Y_{1}^{0}(\hat{k}^{\mathrm{L}})\cos^{2}\beta\vert\vec{D}^{\mathrm{L}}\cdot\vec{F}^{\mathrm{L}}\vert^{2}\label{eq:b10_2_lin_circ}
\end{equation}

where we have added a prime in order to distinguish it from the $b_{1,0}^{\left(2\right)}$
coefficient in Eq. (\ref{eq:b10_2}), and we have $\vec{F}_{1}^{\mathrm{L}}=F\left(0,0,1\right)$
and $\vec{F}_{2}^{\mathrm{L}}=F\left(1,i,0\right)/\sqrt{2}$. Using
Eqs. (\ref{eq:b_lm_2_circ}) and (\ref{eq:b10_2_lin_circ}) we obtain

\begin{align}
2b_{1,0}^{\left(2\right)}+b_{1,0}^{\prime\left(2\right)}\left(k\right) & =\vert A^{\left(2\right)}\vert^{2}d^{2}\left|F\right|^{2}\int\mathrm{d}\Omega_{k}^{\mathrm{M}}\int\mathrm{d}\varrho\,Y_{1}^{0}(\hat{k}^{\mathrm{L}})\vert\vec{D}^{\mathrm{L}}\cdot\vec{F}_{2}^{\mathrm{L}}\vert^{2}\nonumber \\
 & =\vert A^{\left(2\right)}\vert^{2}d^{2}\left|F\right|^{4}\left\{ \int\mathrm{d}\Omega_{k}^{\mathrm{M}}\,\hat{k}^{\mathrm{M}}\cdot\vec{B}^{\mathrm{M}}\right\} \sigma\nonumber \\
 & =\frac{C\sigma}{2\sqrt{3}}\mathcal{B}_{0,0}^{Y}
\end{align}

where in the second line we solved the integral over orientations
as in Eq. (\ref{eq:b10_appendix}) and in the third line we used $\vec{Y}_{0,0}(\hat{k})=\hat{k}/\sqrt{4\pi}$,
Eq. (\ref{eq:B_expansion-1}) and the orthonormality of the spherical
harmonics. Using Eq. (\ref{eq:b10_2_VSH}) for $b_{1,0}^{\left(2\right)}$
yields Eq. (\ref{eq:b10_2_lin}).

\bibliographystyle{apsrev4-1}
\bibliography{MyLibrary}

\end{document}